\documentclass[format=acmsmall, review=false, screen=true]{acmart}

\renewcommand\footnotetextcopyrightpermission[1]{}

\usepackage{multirow}
\usepackage{amssymb,amsmath}
\usepackage{graphicx}
\usepackage{subfig}
\usepackage{algorithm}
\usepackage{algorithmic}
\usepackage{color, soul}
\usepackage{array}
\usepackage{cases}
\usepackage{url}
\usepackage{booktabs}

\usepackage{colortbl}
\definecolor{Gray}{gray}{0.9}

\newcommand{\sampling}{\texttt{sampling} }

\newcounter{MANumberOfComments}
\stepcounter{MANumberOfComments}

\newcounter{LINumberOfComments}
\stepcounter{LINumberOfComments}

\begin{document}

\title{Probabilistic Synchronous Parallel}

\author{Liang Wang}
\orcid{}
\affiliation{%
  \institution{Computer Laboratory, University of Cambridge}
  \city{Cambridge}
  \postcode{CB3 0FD}
  \country{UK}}
  
\author{Ben Catterall}
\orcid{}
\affiliation{%
  \institution{Computer Laboratory, University of Cambridge}
  \city{Cambridge}
  \postcode{CB3 0FD}
  \country{UK}}

\author{Richard Mortier}
\orcid{}
\affiliation{%
  \institution{Computer Laboratory, University of Cambridge}
  \city{Cambridge}
  \postcode{CB3 0FD}
  \country{UK}}

\begin{abstract}
Most machine learning and deep neural network algorithms rely on certain iterative algorithms to optimise their utility/cost functions, e.g. Stochastic Gradient Descent (SGD). In distributed learning, the networked nodes have to work collaboratively to update the model parameters, and the way how they proceed is referred to as synchronous parallel design (or barrier control). Synchronous parallel protocol is practically the building block of all distributed learning frameworks, and its design has direct impact on the performance and scalability of the system.

In this paper, we propose a new barrier control technique - Probabilistic Synchronous Parallel (PSP). Comparing to the previous Bulk Synchronous Parallel (BSP), Stale Synchronous Parallel (SSP), and (Asynchronous Parallel) ASP, the proposed solution effectively improves both the convergence speed and the scalability of the SGD algorithm by introducing a sampling primitive into the system. Moreover, we also show that the sampling primitive can be composed with the existing barrier control mechanisms to derive fully distributed PSP-based synchronous parallel.

We not only provide a thorough theoretical analysis\footnote{Most of the theoretical analysis was part of Ben Catterall's master thesis on his part III project,in the Computer Lab, at the University of Cambridge, in 2017.} on the convergence guarantee of PSP-based SGD algorithm, but also implement a full-featured distributed learning framework called Actor System and perform intensive evaluation atop of it.
\end{abstract}

\maketitle

\section{Introduction}
\label{sec:intro}


Barrier synchronisation is critical in many distributed machine learning algorithms. In general, there are three major ways to coordinate how the nodes in a system should progress in iterative learning algorithms: Bulk synchronous parallel (BSP) \cite{Valiant:1990:BMP:79173.79181}, Stale synchronous parallel (SSP) \cite{NIPS2011_4247, NIPS2013_4894, tseng1991rate}, and Asynchronous parallel (ASP) \cite{Low:2012:DGF:2212351.2212354}.

Even though these synchronisation methods have attracted a lot of attentions lately in the distributed machine learning community, they have recurred several times in the distributed computing literature in the past decades. Among the aforementioned three methods, BSP is the most strict one and requires all the nodes making progress in a lockstepped way. BSP is the default option in many map-reduce-based applications. On the other hand, ASP removes such strict synchronisation requirement completely hence nodes can advance their computations without coordinating with other nodes. SSP is a solution in the middle of aforementioned two extremes wherein a bounded delay is specified between the fastest node and slowest one in the system.

Regarding the pros and cons of each solution, BSP is deterministic and can lead to the same machine learning algorithm design. However, the network and system layer have to implement much more complicated logic to deal with network dynamics (nodes can fail and their progress may not be consistent). ASP eases the communication and synchronisation design among the nodes, but introduces errors (often non-negligible) when updating model parameters in the learning algorithms hence the convergence is not always guaranteed. SSP on the other hand tries to make a trade-off between the efficiency (from ASP) and the accuracy (from BSP) by specifying a bounded staleness.

Most literature in the recent years have been investigating the design of synchronisation parallel in a datacenter context, wherein a very stable network environment is assumed. However, as we move into the context where datasets are distributed in a much larger geographical area with less reliable network and inevitable churns (nodes join and leave). There is an obvious question confronting us - "Are these current solutions sufficient (regarding scalability, complexity, accuracy, and etc.) in such a context?" In other words, is the deterministic way of monitoring and controlling the synchronisation barrier really suitable in a highly dynamic environment?

The key design decisions we must take into account in different context are: 

\begin{enumerate}

\item Tight synchronisation among the nodes in order to maximise the value (i.e. accuracy) of each parameter update, but at the price of slower iteration rate.

\item Fast iteration rate with relaxed synchronisation requirement at the price of noisy parameter update, which has negative impact on the convergence rate.

\end{enumerate}

In this paper, we study the design of barrier control mechanism in a much larger and more dynamic system (refer to Section \ref{sec:model} for details). We first compare the existing solutions in the specified context. Then we propose a solution in which nodes are organised in a structured overlay, the barrier is controlled in a probabilistic way, i.e., by sampling the nodes participated in learning to estimate the percentage of the system having finished the current step.

In current systems, model consistency and barrier control are tightly coupled, namely there is often one (logical) server assigned to update model parameters and coordinate the progress of the nodes in an iterative learning algorithm. We show that by introducing a system primitive "sampling", we can decouple the barrier control from model consistency. Moreover, the proposed sampling primitive can be used to compose with the existing BSP and SSP to construct "higher-order" barrier control which further leads to a more scalable fully distributed solution. 

As we mentioned, SSP is designed to provide a tunable parameter between SSP and BSP, in order to balance between the speed of iteration and accuracy of updates in each iteration. However, from system perspective, SSP still requires a global knowledge of the whole network which in the end does not save any communication cost at all. On the other hand, PSP is able to provide the same effect with much less communication overhead, avoiding throttling the server while the system grows much bigger. Moreover, PSP is ignorant of the fact whether it is a centralised or distributed solution, indicating a fully distributed barrier control can be implemented on top of it. PSP allows us to tune between an expensive (homogeneous datacenter) paradigm (with the benefits of tight synchronisation and less noise) to cheap (heterogeneous edge computing) paradigm (with the benefits of loose synchronisation control, fast iteration but more noise).

More specifically, our main contributions are as follows.

\begin{itemize}

\item We propose adding a new system primitive \sampling to modern data analytical platforms, in order to enhance the performance and scalability of current iterative learning algorithms, especially in a heterogeneous and dynamic network.

\item The sampling primitive can be composed with the existing barrier controls like BSP and SSP to derive fully distributed synchronisation parallel.

\item We have implemented the full system and evaluated the proposed solution Probabilistic Synchronise Parallel (PSP) at a large scale. We show that PSP achieves better trade-off than existing solutions, i.e., iterate faster than SSP and with more accurate update.

\item We perform a through theoretical analysis over PSP, and our results show that PSP is able to provide probabilistic convergence guarantee (as a function of sample size). Both evaluation and theoretical results indicate that a small sample size is already able to bring most of the benefits of PSP. 

\end{itemize}

\section{Background}
\label{sec:background}

Table \ref{table:barriercontrolmethodclassification} summarises the synchronisation control used in different machine learning systems. BSP appears to be a more popular choice due to its determinism nature. The proposed Actor system has implemented all the existing barrier controls.

\begin{table}[h!]
\centering
\begin{tabular}{|c|c|c|}
\hline
\textbf{System} & \textbf{Synchronisation} & \textbf{Barrier Method} \\

\hline
MapReduce~\cite{dean2004} & Requires map to complete before reducing &  BSP\\
\hline
Spark~\cite{spark} & Aggregate updates after task completion & BSP \\
\hline
Pregel~\cite{pregel} & Superstep model & BSP \\
\hline
Hogwild!~\cite{hogwild} & ASP but system-level bounds on delays & ASP, SSP \\
\hline 
Parameter Servers~\cite{LiParamServer} & Swappable synchronisation method & BSP, ASP, SSP \\
\hline 
Cyclic Delay~\cite{langford2009} & Updates delayed by up to $N-1$ steps  & SSP \\
\hline
 Yahoo! LDA~\cite{yahoolda} & Checkpoints &  SSP, ASP \\
\hline
 Owl+Actor\cite{liang2017owl} & Swappable synchronisation method & BSP, ASP, SSP, PSP \\
\hline
\end{tabular}
\caption{Classification of the synchronisation methods used by different systems.}
\label{table:barriercontrolmethodclassification}
\end{table}

\paragraph{Bounded Synchronous Parallel (BSP)} BSP is a deterministic scheme where workers  perform a computation phase followed by a synchronisation/communication phase where they exchange updates~\cite{Xing2016}. The method ensures that all workers are on the same iteration of a computation by preventing any worker from proceeding to the next step until all can. Furthermore, the effects of the current computation are not made visible to other workers until the barrier has been passed. Provided the data and model of a distributed algorithm have been suitably scheduled, BSP programs are often serializable --- that is, they are equivalent to sequential computations. This means that the correctness guarantees of the serial program are often realisable making BSP the strongest barrier control method~\cite{ho2013}.  Unfortunately, BSP does have a disadvantage. As workers must wait for others to finish, the presence of \textit{stragglers}, workers which require more time to complete a step due to random and unpredictable factors~\cite{Xing2016}, limit the computation efficiency to that of the slowest machine. This leads to a dramatic reduction in performance. Overall, BSP tends to offer high computation accuracy but suffers from poor efficiency in unfavourable environments.

\paragraph{Asynchronous Parallel (ASP)}  ASP takes the opposite approach to BSP, allowing computations to execute as fast as possible by running workers completely asynchronously. In homogeneous environments (e.g. datacenters), wherein the workers have similar configurations, ASP enables fast convergence because it permits the highest iteration throughputs. Typically, $P$-fold speed-ups can be achieved~\cite{Xing2016} by adding more computation/storage/bandwidth resources. However, such asynchrony causes delayed updates: updates calculated on an old model state which should have been applied earlier but were not. Applying them introduces noise and error into the computation. Consequently, ASP suffers from decreased iteration quality and may even diverge in unfavourable environments. Overall, ASP offers excellent speed-ups in convergence but has a greater risk of diverging especially in a heterogeneous context.

\paragraph{Stale Synchronous Parallel (SSP)} SSP is a bounded-asynchronous model which can be viewed as a relaxation of BSP. Rather than requiring all workers to be on the same iteration, the system decides if a worker may proceed based on how far behind the slowest worker is, i.e. a pre-defined bounded staleness. Specifically, a worker which is more than $s$ iterations behind the fastest worker is considered too slow. If such a worker is present, the system pauses faster workers until the straggler catches up. This $s$ is known as the \textit{staleness} parameter.

More formally, each machine keeps an iteration counter, $c$, which it updates whenever it completes an iteration. Each worker also maintains a local view of the model state. After each iteration, a worker commits updates, i.e., $\Delta$, which the system then sends to other workers, along with the worker's updated counter. The bounding of clock differences through the staleness parameter means that the local model cannot contain updates older than $c -s - 1$ iterations. This limits the potential error. Note that systems typically enforce a \textit{read-my-writes} consistency model.

The staleness parameter allows SSP to provide deterministic convergence guarantees~\cite{Xing2016,dai2014,ho2013}. Note that SSP is a generalisation of BSP: setting $s = 0$ yields the BSP method, whilst setting $s = \infty$ produces ASP. Overall, SSP offers a good compromise between fully deterministic BSP and fully asynchronous ASP~\cite{ho2013}, despite the fact that the central server still needs to maintain the global state to guarantee its determinism nature.

\section{System Model}
\label{sec:model}


In contrast to a highly reliable and homogeneous datacenter context, We assume a distributed system consisting of a larger amount of (tens of thousands of) heterogeneous nodes that are distributed at a much larger geographical areas (e.g., many different cities). The network is unreliable since links can break down, and the bandwidth is heterogeneous. The nodes are not static, they can join and leave the system at any time. Therefore, the churn is not negligible.

Each node holds a local dataset. Even though nodes can query each other, we do not assume any specific information sharing between nodes or between a node and a centralised server. For the data analytic applications running atop of the nodes, we focus on the following algorithms: stochastic gradient descent algorithm since it is one of the few core algorithms in many machine learning (ML) and deep neural network (DNN) algorithms.


\subsection{Problem Formulation}
\label{sec:problem}

First, we need to understand why barrier control mechanism is critical to a iterative learning algorithms. What would happen if nodes' barriers are not synchronised? Often, in data parallel computation, the shared model parameters will be updated based on the individual updates from each piece of data. More precisely, the aggregated updates will be used to update the model. Unsynchronised updates will introduce errors into the model. Similarly, for model parallel algorithms, the inaccuracy can also be introduced in the same way if barriers are not aligned.

However, one thing worth noting here is that practically many iterative learning algorithms can tolerate certain level of errors in the process of converging to final solutions. Given a well-defined boundary, if most of the nodes have passed it, the impact of those lagged nodes should be minimised. Therefore, in a very unreliable environment, we can minimise the impact by guaranteeing majority of the system have synchronised boundary. The extent that a barrier is synchronised represents the trade-off between the accuracy (in each iteration) and efficiency (the speed of convergence) of an iterative learning algorithm.

Then the immediate question is how to estimate what percent of nodes have passed a given synchronisation barrier? We need two pieces of information: 

\begin{enumerate}

\item an estimate on the total number of nodes in the system;

\item an estimate of the distribution of current steps of the nodes.

\end{enumerate}

However, as the system grows bigger, monitoring all the nodes in order to maintain the global state centrally eventually becomes infeasible due to communication cost on the control plane. Moreover, setting up a centralised server introduces a typical single point of failure. 

In the following of this paper, we assume a network which can be represented as a directed graph $G = (V, E)$, where $V$ represents the node set and $E$ represents the link set. For a given time $t$, we use $s_{i,t}$ to denote the node $v_i$'s step at time $t$. In some cases, we drop subscript $t$ and simply write $s_i$ if the context is self-explained. For the ease of reading, we do not write down all the notations but choose to introduce them gradually along with our theoretical analysis. 

\subsection{A Potential Solution}

To obtain the aforementioned two pieces of information, we can organise the nodes into a structured overlay (e.g., chord\cite{Stoica:2001:CSP:383059.383071} or kademlia\cite{Maymounkov2002}), the total number of nodes can be estimated by the density of each zone (i.e., a chunk of the name space with well-defined prefixes), given the node identifiers are uniformly distributed in the name space. Using a structured overlay in the design guarantees the following sampling process is correct, i.e., random sampling. 


We then (randomly) select a subset of nodes in the system and query their individual current local step. By so doing, we can obtain a sample of the current nodes' steps in the whole system. By investigating the distribution of these observed steps, we can derive an estimate of the percentage of nodes which have passed a given step. 




After deriving the estimate on the step distribution, a node can choose to either pass the barrier by advancing its local step if a given threshold has been reached (with certain probability) or simply holds until certain condition is satisfied.


\section{System Architecture}
\label{sec:architecture}

We have developed a distributed data processing framework called Actor to demonstrate our proposal. The system has implemented core APIs in both map-reduce\cite{Dean:2010:MFD:1629175.1629198} engine and parameter sever\cite{186214} engine. Both map-reduce and parameter server engines need a (logical) centralised entity to coordinate all the nodes' progress.  To demonstrate PSP's capability to transform an existing barrier control method into its fully distributed version, we also extended the parameter server engine to peer-to-peer (p2p) engine. The p2p engine can be used to implement both data and model parallel applications, both data and model parameters can be (although not necessarily) divided into multiple parts then distributed over different nodes. 

Each engine has its own set of APIs. E.g., map-reduce engine includes \texttt{map}, \texttt{reduce}, \texttt{join}, \texttt{collect}, and etc.; whilst the peer-to-peer engine provides four major APIs: \texttt{push}, \texttt{pull}, \texttt{schedule}, \texttt{barrier}. It is worth noting there is one function shared by all the engines, i.e. \texttt{barrier} function which implements various barrier control mechanisms.

As an example, we briefly introduce the interfaces in peer-to-peer engine.

\begin{itemize}

\item \texttt{schedule}: decide what model parameters should be computed to update in this step. It can be either a local decision or a central decision.

\item \texttt{pull}: retrieve the updates of model parameters from somewhere then applies them to the local model. Furthermore, the local updates will be computed based on the scheduled model parameter.

\item \texttt{push}: send the updates to the model plane. The updates can be sent to either a central server or to individual nodes depending on which engine is used(e.g., map-reduce, parameter server, or peer-to-peer). 

\item \texttt{barrier}: decide whether to advance the local step. Various synchronisation methods can be implemented. Besides the classic BSP, SSP, and ASP, we also implement the proposed PSP within this interface.

\end{itemize}



\subsection{Possible Design Combinations}

Practically all iterative learning algorithms are stateful. Both model and nodes' states need to be stored somewhere in order to coordinate nodes to make progress in a training process. Regarding the storage location of the model and nodes' states, there are four possible combinations as below (\textbf{states} in the list refer to the nodes' states specifically):

\begin{enumerate}

\item \textbf{[centralised model, centralised states]}: the central server is responsible for both synchronising the barrier and updating the model parameters. E.g., Map-reduce and parameter server fall into this category.

\item \textbf{[centralised model, distributed states]}:  the central server is responsible for updating the model only. The nodes coordinate by themselves to synchronise the barrier (in a distributed way). P2P engine falls into this category.

\item \textbf{[distributed model, centralised states]}: this combination in practice is rare because it is hard to justify its actual benefits.

\item \textbf{[distributed model, distributed states]}: both updating model and synchronising barrier is performed in a distributed fashion. A model can be divided into multiple chunks and distributed among different nodes.

\end{enumerate}

All BSP, ASP, SSP, PSP can be used in case 1; only ASP and PSP can be used in case 2 and 4. Case 3 is ignored at the moment. With PSP, the sever for maintaining the model can become "stateless" since it does not have to possess the global knowledge of the network.  For the server in case 2 especially, its role becomes a stream server which continuously receives and dispatches model updates. This can significantly simplify the design of various system components.

Often, model update and barrier synchronisation are tightly coupled in most previous design. However, by decoupling these two components with sampling primitive, we can achieve better scalability with reasonable convergence degradation.

\subsection{Compatibility with Existing Synchronisation Primitives}



\begin{algorithm}[!tb]
  \caption{\texttt{barrier} function for classic BSP}
  \label{alg:bsp}
  \begin{algorithmic}[1]
    \STATE{\textbf{Input:}}
    \STATE{\qquad Global state of all nodes $V$}
    \STATE{\textbf{Output:}}
    \STATE{\qquad $step_i = step_j$ $(\forall v_i, v_j \in V)$}
  \end{algorithmic}
\end{algorithm}

\begin{algorithm}[!tb]
  \caption{\texttt{barrier} function for classic SSP}
  \label{alg:ssp}
  \begin{algorithmic}[1]
    \STATE{\textbf{Input:}}
    \STATE{\qquad Global state of all nodes $V$}
    \STATE{\qquad Staleness $\theta$}
    \STATE{\textbf{Output:}}
    \STATE{\qquad $| step_i - step_j | \leq \theta$ $(\forall v_i, v_j \in V)$}
  \end{algorithmic}
\end{algorithm}

Algorithm~\ref{alg:bsp} and \ref{alg:ssp} present the high-level logic of classic BSP and SSP that can be implemented in the \texttt{barrier} function. The \texttt{barrier} function is called by the centralised server to check the synchronisation condition with the given inputs. The output of the function is a boolean decision variable on whether or not to cross the synchronisation barrier, depending on whether the criterion specified in the algorithm is met.

With the proposed \sampling primitive, almost nothing needs to be changed in aforementioned algorithms except that only the sampled states instead of the global states are passed into the \texttt{barrier} function. Therefore, we can easily derive the probabilistic version of BSP and SSP, namely pBSP and pSSP. 

However, it is worth emphasising that applying \sampling leads to the biggest difference between the classic synchronisation control and probabilistic control: namely the original synchronisation control requires a centralised node to hold the global state whereas the derived probabilistic ones no longer require such information thus can be executed independently on each individual node, further leading to a fully distributed solution.




\section{Preliminary Evaluation}
\label{sec:evaluation}

In the evaluation, we aim to answer the following questions.

\begin{enumerate}

\item Can PSP achieve faster iteration and better accuracy than SSP, i.e. faster convergence? 


\item Comparing to ASP, which has the fast iteration rate, can PSP achieve better accuracy?

\item How does different synchronisation control impact the distribution of lags? 

\item How does the number of control messages grow as the system grows in each solution? 


\end{enumerate}


To simplify the evaluation, we can assume that every node hold the equal-size data and the data is \textit{i.i.d}. 
We consider both centralised and distributed scenarios while evaluating PSP. In the centralised scenario, the central server applies \sampling primitive and the PSP is as trivial as a counting process , because the central server has the global knowledge of the states of all nodes. In the distributed scenario, each individual node performs \sampling locally whenever they need to make the decision on crossing a barrier. We use Owl\cite{liang2017owl} library for all the numerical functions needed in the evaluation.

\subsection{Impacts on System Performance}

In the following experiments, each node takes a sample of 1\% (e.g., 10 nodes given a 1000 node network) of the system size, unless otherwise specified.

\begin{figure}[!htp]
  \centering 
  \subfloat[Progress distribution in steps]{\label{fig:barrier:1}\includegraphics[width=8cm]{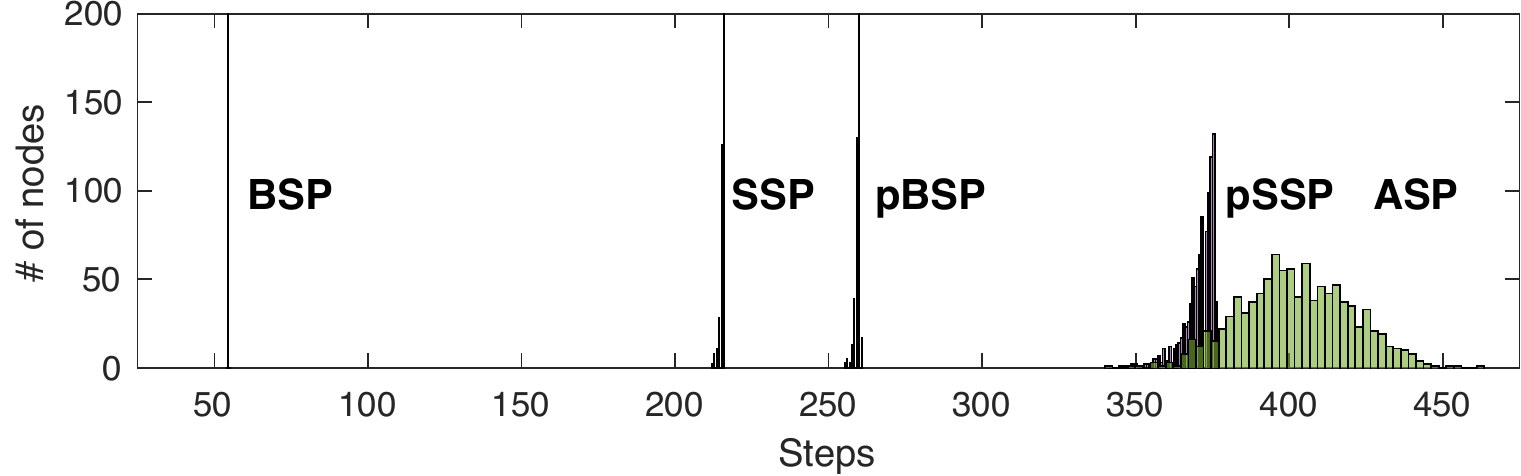}}\\
  \subfloat[CDF of nodes as a function of progress. Algorithm behaviour is controlled by the single parameter sample size to mimic other solutions. No global state is maintained by any single node.]{\label{fig:barrier:4}\includegraphics[width=8cm]{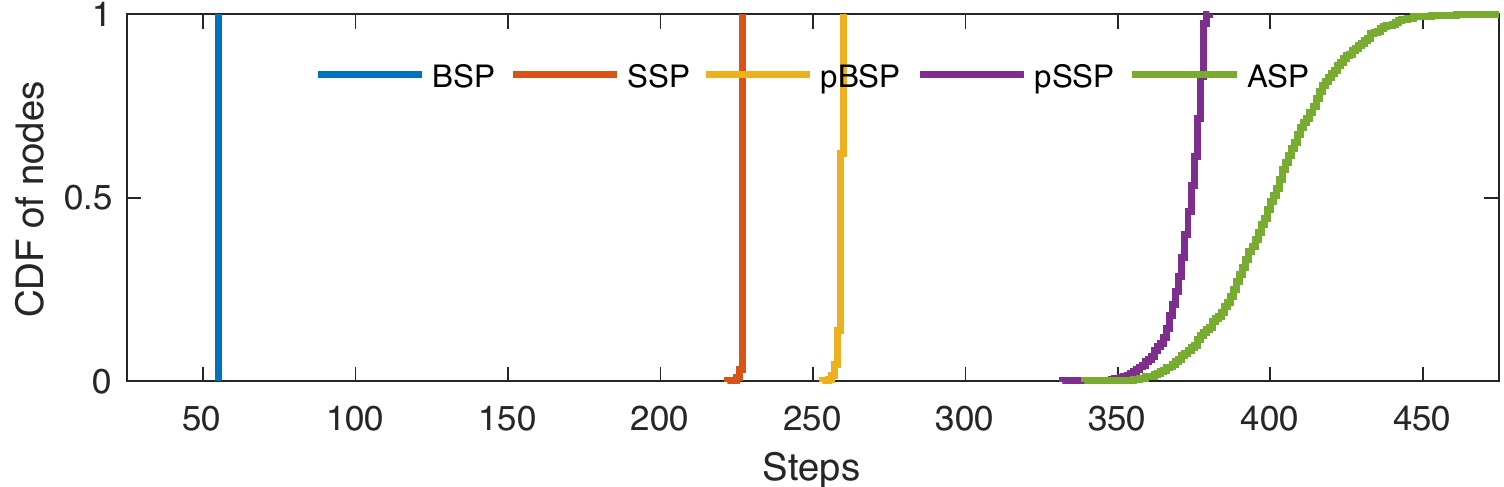}}\\
  \subfloat[pBSP parameterised by different sample sizes, from 0 to 64. Increasing the sample size make the curves shift from right to left with decreasing spread, covering the whole spectrum from the most lenient ASP to the most strict BSP.]{\label{fig:barrier:5}\includegraphics[width=8cm]{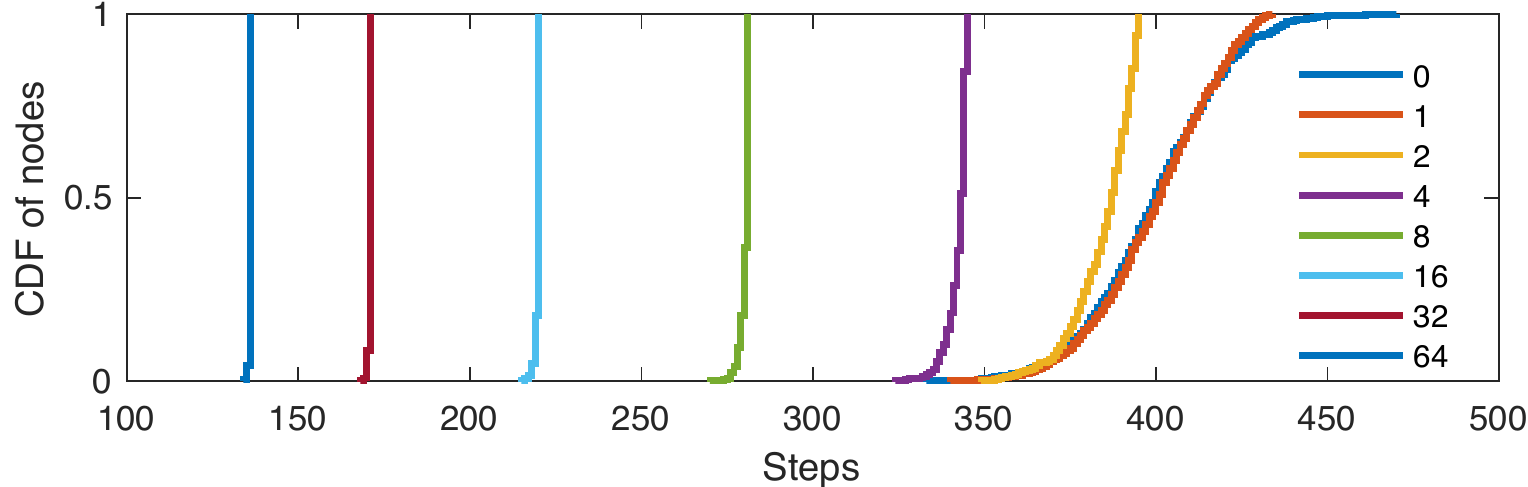}}\\
  \subfloat[Normalized error value; note that the measurements all taken at the same times (5\,s, 10\,s, etc.)]{\label{fig:barrier:2}\includegraphics[width=8cm]{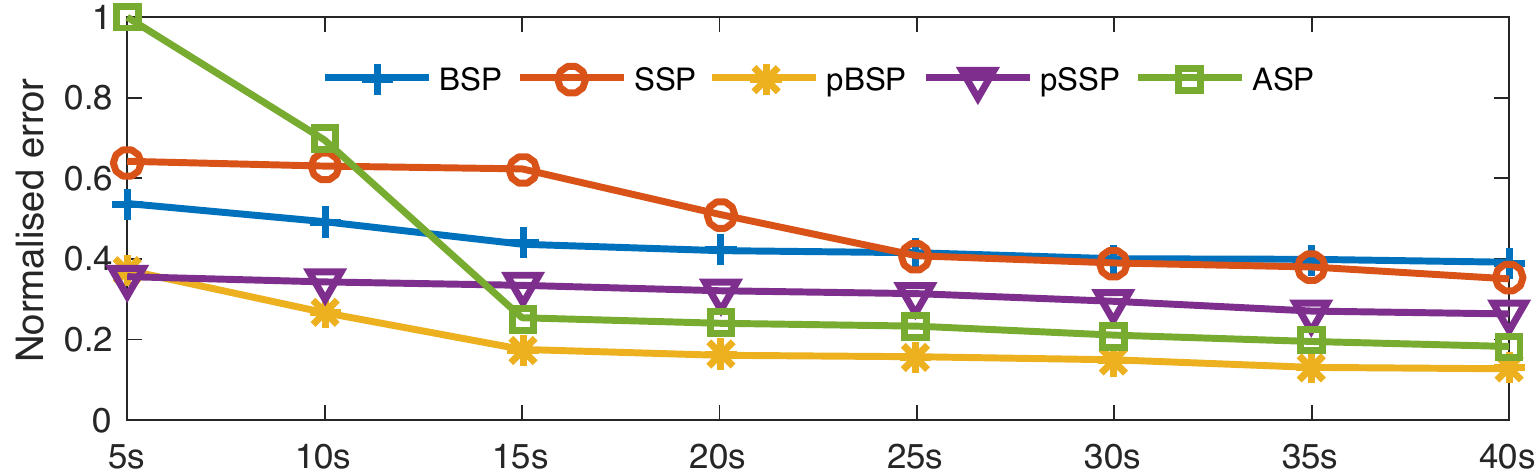}}\\
  \subfloat[Aggregated number of updates received by the server]{\label{fig:barrier:3}\includegraphics[width=8cm]{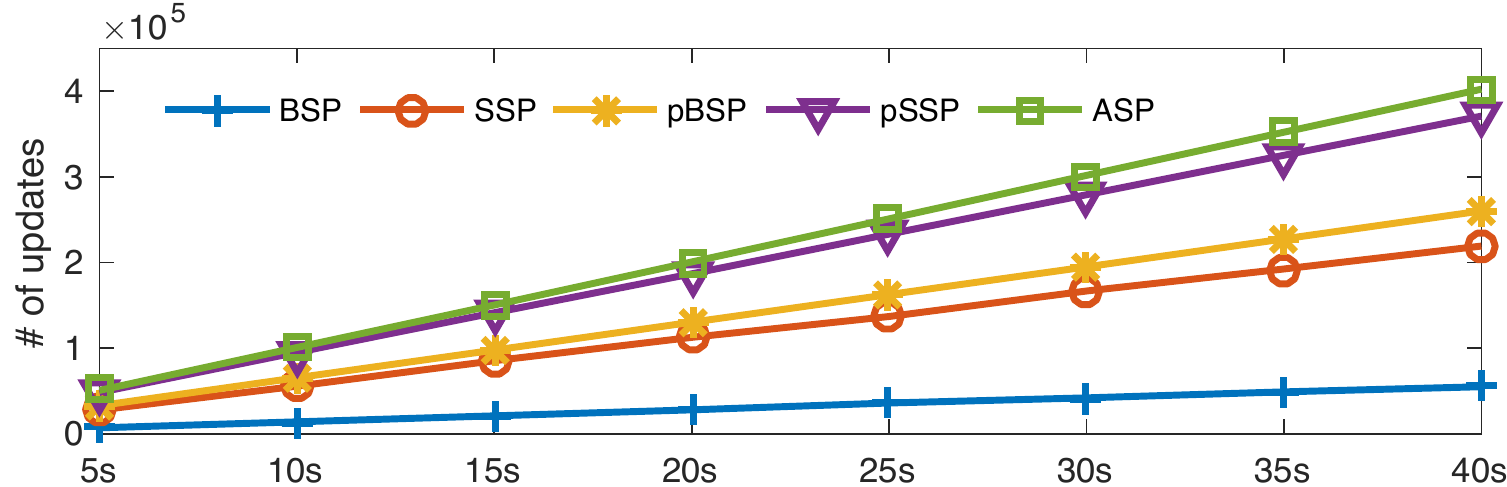}}
  \caption{Comparison of SGD (stochastic gradient descendent) using five different barrier control strategies, we can see that probabilistic synchronous parallel achieves good trade-off between efficiency and accuracy.}
  \label{fig:barrier}
\end{figure}

Fig.\ref{fig:barrier} shows the results of evaluating PSP by simulating five different barrier control strategies for 40\,seconds on a network of 1000 nodes running SGD algorithm. We use the parameter server engine to learn a linear model of 1000 parameters. Fig.\ref{fig:barrier:1} plots the progress in steps of all nodes after the 40 simulated seconds. As expected, the most strict BSP leads to the slowest but most tightly clustered step distribution, while ASP is the fastest but most spread due to no synchronisation at all. SSP allows certain staleness (4 in our experiment) and sits between BSP and ASP. pBSP and pSSP are the probabilistic versions of BSP and SSP respectively, and further improve the iteration efficiency while limiting dispersion.

For the same experiment, Fig.\ref{fig:barrier:2} plots the CDF of nodes as a function of their progress in steps. ASP has the widest spread due to its unsynchronised nature. Fig.\ref{fig:barrier:5} focuses on pBSP synchronisation control with various parametrisation. In the experiment, we vary the sample size from 0 to 64. As we increase the sample size step by step, the curves start shifting from right to the left with tighter and tighter spread, indicating less variance in nodes' progress. With sample size 0, the pBSP exhibits exactly the same behaviour as that of ASP; with increased sample size, pBSP starts becoming more similar to SSP and BSP with tighter requirements on synchronisation. pBSP of sample size 16 behaves very close to SSP regarding its progress rate. 

Another interesting thing we noticed in the experiment is that, with a very small sample size of one or two (i.e., very small communication cost on each individual node), pBSP can already effectively synchronise most of the nodes comparing to ASP. The tail caused by stragglers can be further trimmed by using larger sample size. \textit{This observation is further confirmed by our theoretical analysis later in Section \ref{sec:theory}, which explicitly shows that a small sample size can effectively push the probabilistic convergence guarantee to its optimum even for a large system size, which further indicates the superior scalability of the proposed solution.}


These results indicate that the single parameter pBSP built atop of \sampling primitive already allows us to cover the whole spectrum of synchronisation controls from the most strict BSP to the least one ASP, (more importantly) without requiring any single node to maintain the global state. Even more attractive characteristic of the proposed PSP is that the probabilistic control is able to achieve much higher model accuracy given the same iteration rate, as we will show below.

\subsection{Impacts on Model Accuracy}

Fig.\ref{fig:barrier:2} shows the normalized error, defined as the $L_2$ norm of the difference between the current prediction and the true values of all parameters, of the five strategies. ASP has high initial error but reduces rapidly as it can iterate most quickly, and SSP sits between BSP and ASP as expected. The probabilistic controls (pBSP, pSSP) both allow quick iteration while controlling dispersion, effectively preventing any node becoming too inaccurate. In all cases, pBSP achieves the best accuracy.

Finally, Fig.\ref{fig:barrier:3} shows the number of updates received by the node holding the model (the server) as model training progresses. We only consider a count of messages as message sizes will vary between different learning algorithms, but we ignore control messages among the nodes as their size is negligible compared to the size of model updates in any realistic training. BSP's strict synchronisation requirement means it advances slowly, resulting in comparatively low communication costs, while the absence of synchronization in ASP results in very quick progress and a correspondingly large number of messages -- almost 10x communication overhead compared to BSP. pBSP sits in the middle in terms of communication overhead.

Reading Fig.\ref{fig:barrier:3} in parallel with Fig.\ref{fig:barrier:2} it is clear that there is a trade-off between convergence speed and communication overhead. An algorithm with loosely controlled synchronization barrier will move more quickly but produces less useful updates at each step in terms of the error reduction in the model at the server. The challenge is to balance between convergence speed and communication overhead to achieve an accurate model with little communication cost. pBSP achieves this goal quite well, and it can be further tuned by adjusting the sample size used.




\subsection{Robustness to Stragglers}



\begin{figure*}[!htp]
  \centering 
  \subfloat[Normalised average speed as a function of percentage of the slow nodes from 0\% to 30\%.]{\label{fig:straggler:1:1}\includegraphics[width=4.4cm]{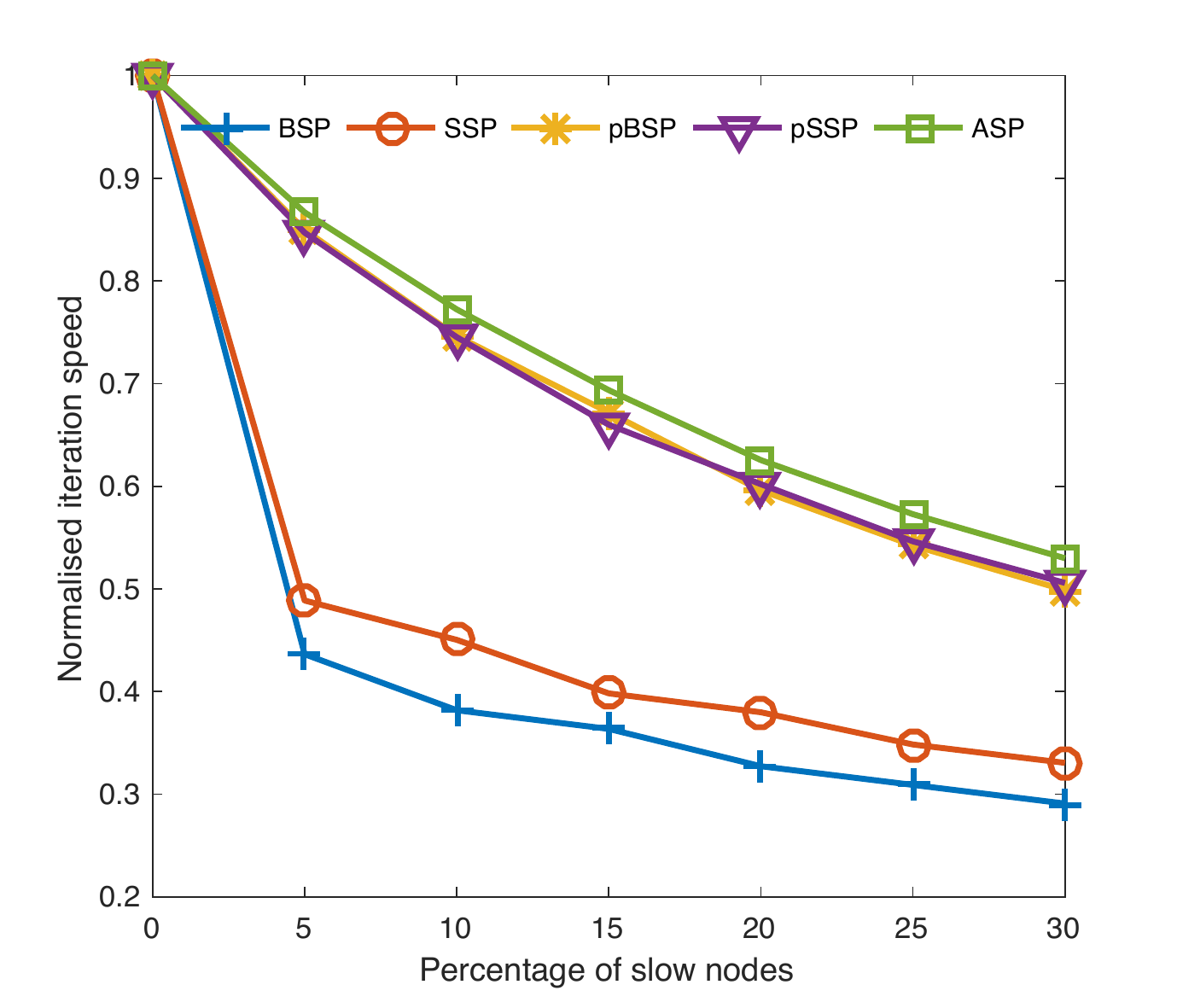}}
  \quad
  \subfloat[Percentage of increased error as a function of percentage of slow nodes from 0\% to 30\%.]{\label{fig:straggler:1:2}\includegraphics[width=4.4cm]{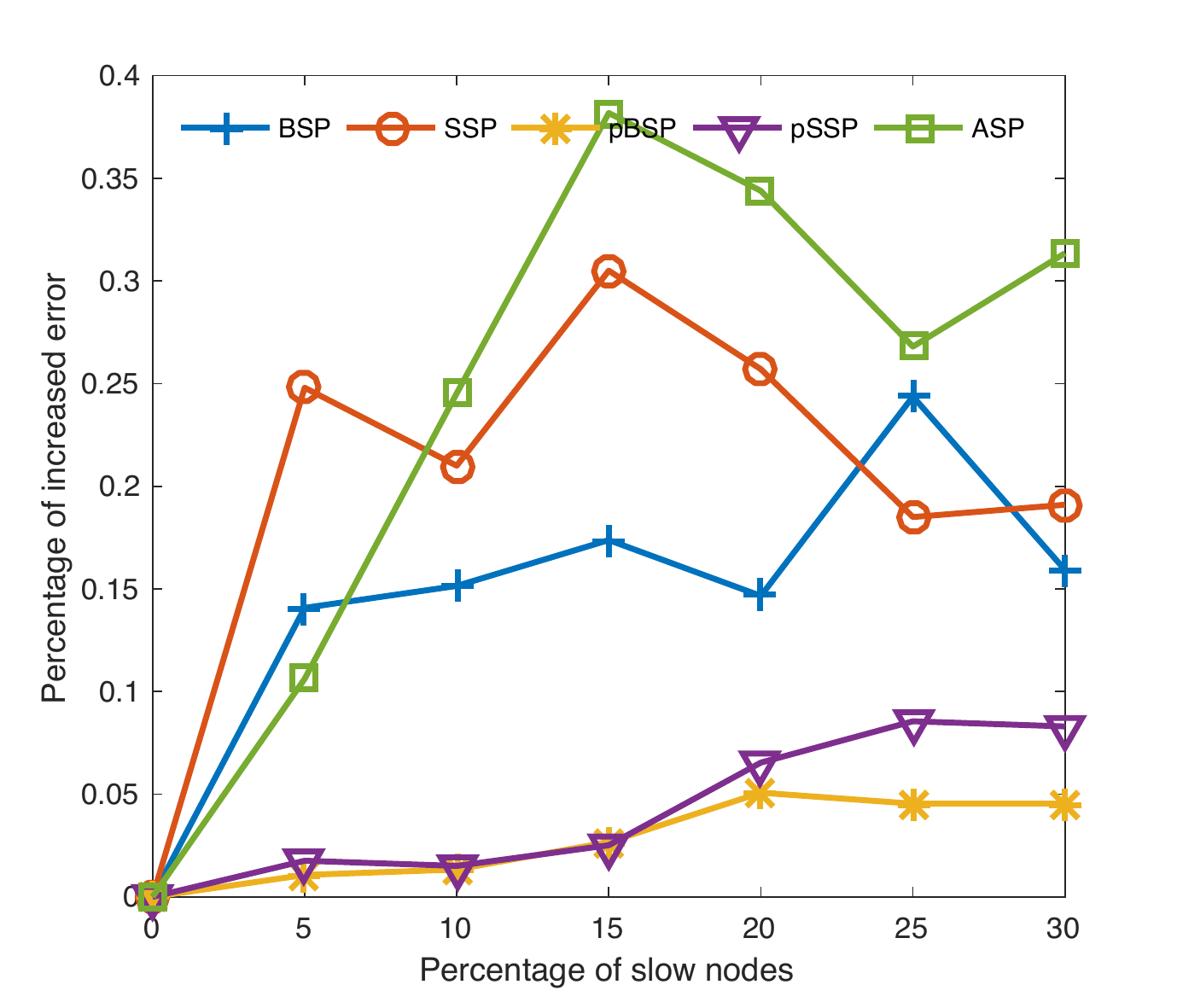}}
  \quad
  \subfloat[Keep 5\% slow nodes, increase their slowness step by step from 2x to 16x slow.]{\label{fig:straggler:1:3}\includegraphics[width=4.4cm]{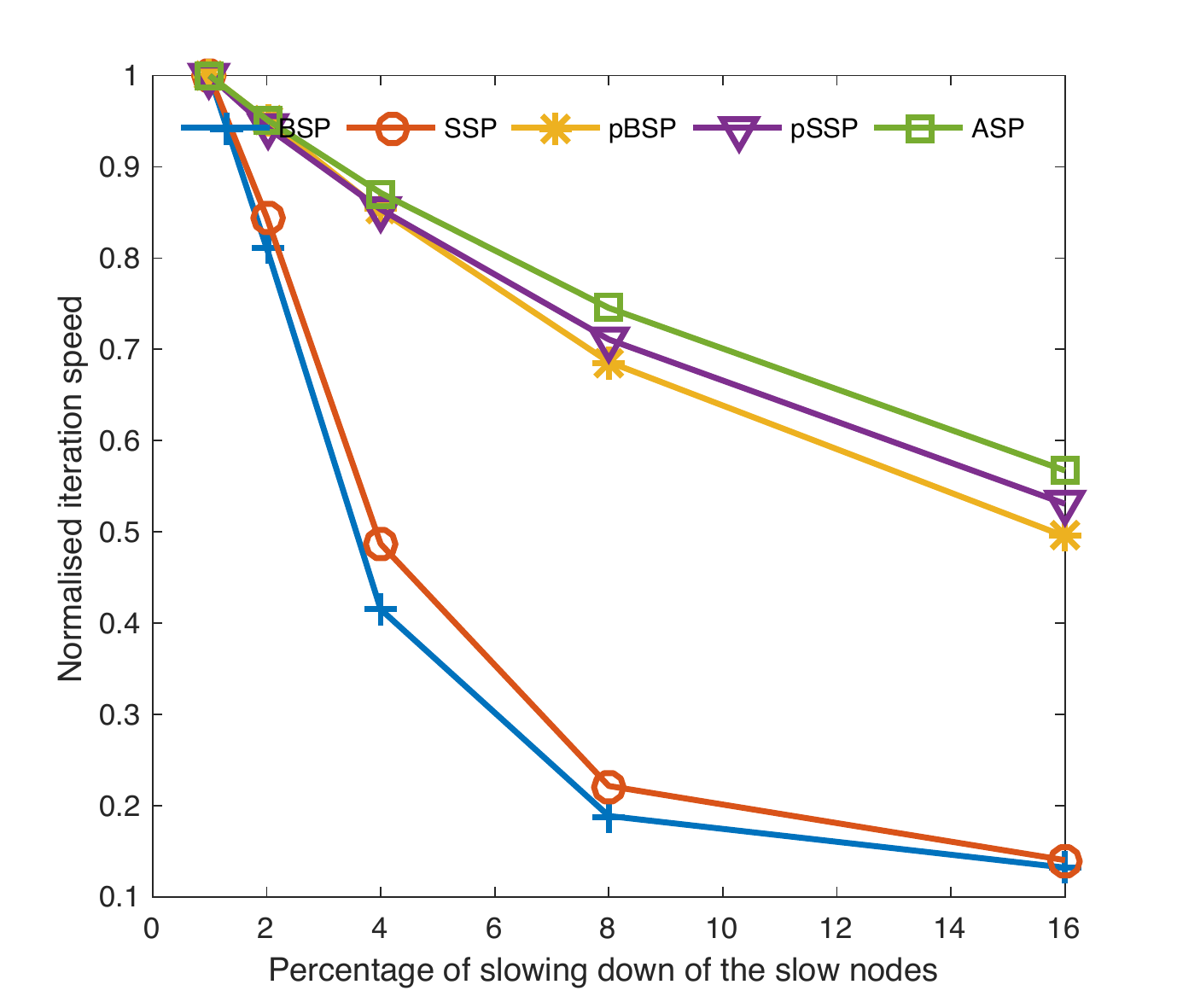}}
  \caption{Stragglers impact both system performance and accuracy of model updates. Probabilistic synchronisation control by sampling primitive is able to mitigate such impacts.}
  \label{fig:straggler:1}
\end{figure*}

In the experiments presented in Figure~\ref{fig:straggler:1}, we study how stragglers can impact both system performance and model accuracy and how PSP can help in mitigating such negative effects. 

In the experiments presented in Figure~\ref{fig:straggler:1:1}, we inject a certain percentage of slow nodes into the system. The slow node is 4x slower than the normal nodes, namely on average they spend 4 times as much time as normal node to finish one iteration. We increase the percentage of the slow nodes step by step from 0\% to 30\%, then we measure the average progress at 40s and calculate the ratio between the systems with and without stragglers (i.e., 0\% case).

As we can see, both BSP and SSP are sensitive to the stragglers, as long as there are some stragglers in the system, both synchronisation strategies slow down the progress significantly. SSP performs slightly better than BSP since it allows certain amount of staleness. On the other hand, with the \sampling primitive, pBSP and pSSP is very similar to ASP, the degradation is close to sub-linear. This is understandable, as more and more slow nodes are injected, the system performance will approach to 25\% of the original performance at its infinity (recall the slow nodes are 4x slower). Another thing worth pointing out is that as the system size increases (keep the number of straggler fixed), the ASP, pBSP, and pSSP curves will shift upwards whilst BSP and ASP curves will remain the same, indicating their robustness to stragglers.

Stragglers can also impact the model accuracy. For BSP and SSP, stragglers slow down the progress to delay the convergence; for ASP and others, stragglers may submit outdated updates which introduces noise and destroys previous updates to diverge the learning. Figure~\ref{fig:straggler:1:2} plots the increased error (due to stragglers) as a function of the percentage of stragglers in the system. More precisely, we first measure the model error at a specific given time when there is no stragglers, then we increase the percent of stragglers step by step as before and measure the model error again at the same given time. The figure plots how much such model error will increase. \textit{As we can see, ASP appears to be the most sensitive to stragglers regarding model error even though it is the least sensitive one regarding to progress.} This is due to the asynchronised nature of SSP, the model has received too many inaccurate updates which "washed out" the other faster nodes' efforts. For BSP and ASP, the increased model error can be explained as a consequence of slowed down progress. As we can see in Figure\ref{fig:straggler:1:1}, BSP and SSP become so slow that they cannot achieve the same accuracy simple due to not enough updates.

In another experiment presented in Figure~\ref{fig:straggler:1:3}, we keep 5\% slow nodes as a constant but we increase the "slowness" step by step. We increase the slowness from 1x (i.e., no slowing down) to 16x slower, then plot the progress distribution as a function of slowness. The figure indicates that both BSP and SSP are completely dominated by the stragglers. In other words, a small amount of stragglers is able to highly influence the overall system performance under these two synchronisation controls. Meanwhile, pBSP, pSSP, and ASP are clearly in another group which are less subject to stragglers. However, it is worth emphasising again that pBSP and pSSP are more robust regarding model error than ASP.

\subsection{Scalability to System Size}

\begin{figure}[!htp]
  \centering 
  \subfloat[Percentage of changes in average progress as a function of systme size.]{\label{fig:systemsize:1}\includegraphics[width=6.0cm]{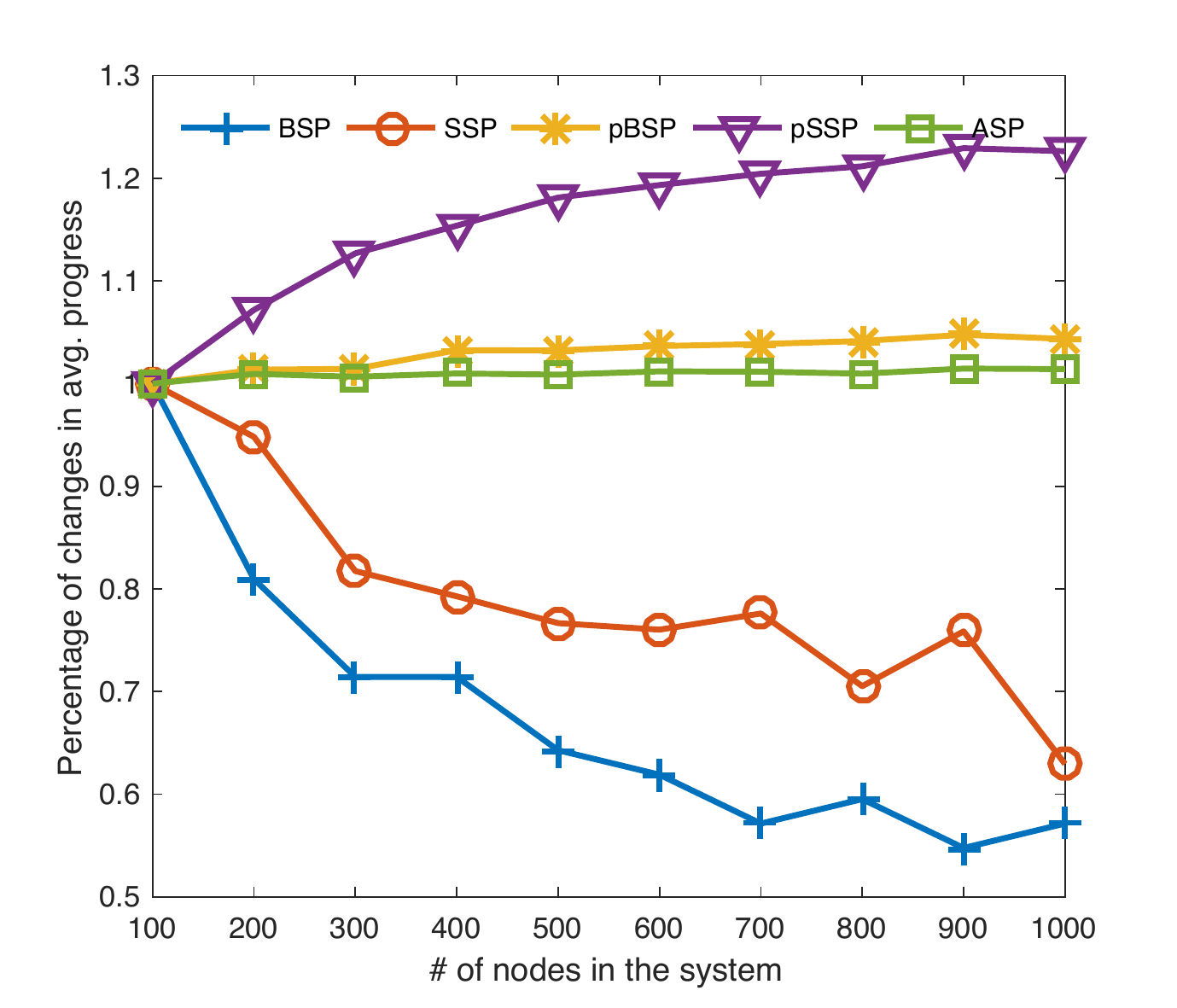}}
  \caption{Probabilistic synchronisation control exhibits strong scalability with increasing system size. In this experiment, a constant of 10-node sample is taken by the nodes. Probabilistic control provides as good scalability as ASP but with much stronger synchronisation guarantees to improve algorithm accuracy.}
  \label{fig:systemsize}
\end{figure}

In Figure~\ref{fig:systemsize}, we present the results wherein we keep 5\% of slow nodes and increase the system size from 100 nodes to 1000 nodes. We set 100-node system as norm, then measure the percentage of changes regarding the average progress at 40s given each system size. According to the results, the performance of both BSP and SSP start dropping as the system size increases, whilst ASP remains constant due to its strategy is independent from the network size. pBSP performs slight better while pSSP actually increases as the system grows (given a fixed sample size), which can be explained as the growing system size "diluted" the impact of stragglers in the sampling process (i.e., the chance of sampling a straggler is reduced). In all cases, probabilistic control exhibits strong scalability thanks to its distributed synchronisation control.

\section{Analysis of Convergence}
\label{sec:theory}

In the following, we present a theoretical analysis of SGD convergence under different synchronisation controls. We first introduce the formal definition of each synchronisation strategy and the proof framework used in our analysis, followed by the detailed convergence proof.

\subsection{Formal Definition of Barrier Control Methods} 
\label{sec:formalbarrier}
To pass a barrier, a worker must satisfy the specified conditions of the barrier control method. Otherwise, the worker needs to wait until the conditions become true. In the following definitions, let $V$ be the set of all workers in the system and $s_{i}$ the step of worker $i$. 
\paragraph{BSP} BSP requires all workers to be on the same step:
\begin{equation}
\forall i,j \in V. s_{i}  = s_{j}\,.
\end{equation}

\paragraph{SSP}
SSP enforces an upper limit on the lag a worker can experience:
\begin{equation}
\forall i,j \in V. |s_{i}  - s_{j}| \leq s\,,
\end{equation} where $s$ is a parameter known as \textit{staleness}.

\paragraph{ASP}
No synchronisation is performed in ASP.
\begin{equation}
\top
\end{equation}

\paragraph{pBSP}
PSP generalises the previous methods. Specifically, only a subset, $S \subseteq V$, of the workers are tested. For pBSP, this yields:
\begin{equation}
\forall i,j \in S \subseteq V . s_{i} = s_{j}\,.
\end{equation}
If $S = V$, then pBSP reduces to BSP and if $S = \emptyset$, then it reduces to ASP.

\paragraph{pSSP}
The most general method of PSP is pSSP. Here, sampling a subset of workers, $S \subseteq V$, leads to:
\begin{equation}
\forall i,j \in S \subseteq V . |s_{i}  - s_{j}| \leq s\,.
\end{equation}
Clearly, if $S = V$, then the method becomes SSP and if $S = \emptyset$ or $s = \infty$, then the method reduces to ASP. Furthermore, if $s = 0$ then pSSP reduces to pBSP. pSSP therefore is the generalisation of all existing barrier control method.

\subsection{General Proof Framework}

To prove convergence, we first consider a general form of a distributed machine learning algorithm. We formalise the process of updating the model by casting this as a sequence of updates. The true sequence of updates encodes the results expected from a fully deterministic barrier control system, such as  BSP. A noisy sequence represents scenarios where updates are reordered due to sporadic and random network and system delays. In the analysis, we consider the different restrictions each barrier control method enforces on this noisy sequence. 

With these notions, the convergence properties of barrier control methods can be analysed. Specifically, we find a bound on the difference between the true sequence and a noisy sequence. By showing that this bound has certain properties, we prove convergence.

As a proof of convergence for ASP requires much of the work needed in a proof for PSP, we first prove that SGD under ASP converges and then refine this proof for PSP.

Dai et al.~\cite{dai2014} presented bounds for SSP demonstrating that, with suitable conditions, SGD under SSP converges deterministically. Ho et al.~\cite{ho2013} presented an alternative method and proved probabilistic convergence for SSP. Our analysis extends the method used for the deterministic bounds but differs when forming the probabilistic bounds.

In distributed machine learning, a series of updates are applied to a stateful model, represented by a $d$-dimensional vector, $\mathbf{x} \in \mathbb{R}^{d}$, at clock times denoted by $c \in \mathbb	{Z}$. Let the number of workers in the system be $P$ and let  $\mathbf{u}_{p,c} \in \mathbb{R}^{d}$  be the $p$th worker's updates at clock time $c$. These updates are applied to the system state, $\mathbf{x}$. Let $t$ be the index of the serialised updates. 

Define the true sequence of updates as taking the sum over all the workers $t \text{ mod } P$ and then over all iterations,  $\left \lfloor \frac{t}{P} \right \rfloor$:
\begin{equation}
\mathbf{x}_{t} = \mathbf{x}_{0} + \sum_{j = 0}^{t} \mathbf{u}_{j}\,,
\end{equation}	
where $\mathbf{u}_{j} = \mathbf{u}_{t \text{ mod } P, \left \lfloor \frac{t}{P} \right \rfloor}$. The sum is taken as SGD aggregates updates by summing them before applying them to the model. The ordering of the sequence is chosen to represent a deterministic BSP-like system, which would aggregate all updates from the workers and then proceed to the next step. It is worth emphasising that multiple values of $t$ occur in a given iteration, $\left \lfloor \frac{t}{P} \right \rfloor$, and clock time, $c$.

As some updates will be in progress in the network, we have a noisy state, $\tilde{\mathbf{x}}_{p,c}$, which is read by worker $p$ at time $c$.  Let $\tilde{\mathbf{x}}_{t} = \tilde{\mathbf{x}}_{t \text{ mod } P, \lfloor \frac{t}{P} \rfloor}$ so that,
\begin{equation} \label{eq:pspupdates}
\tilde{\mathbf{x}}_{t} = \mathbf{x}_{t} - \sum_{i \in \mathcal{A}_{t}} \mathbf{u}_{i} + \sum_{i \in \mathcal{B}_{t}} \mathbf{u}_{i}\,.
\end{equation}
Here, $\mathcal{A}_{t}$ and  $\mathcal{B}_{t}$ are the index sets of updates where  $\mathcal{A}_{t}$ holds missing updates (those missing from the noisy representation but which are in the true sequence) and  $\mathcal{B}_{t}$ holds the extra updates (those in the noisy sequence but which are not in the true sequence).  
 	
To perform convergence analysis, the difference between the true sequence, $\mathbf{x}_{t}$, and the noisy sequence, $\tilde{\mathbf{x}}_{p,c}$, can be examined.

\subsection{Convergence of SGD under ASP}

\subsection*{Theorem 1: SGD under ASP}
Let $f(\mathbf{x}) = \sum_{t=1}^{T} f_{t}(\mathbf{x})$ be a convex function where each $f_{t} \in \mathbb{R}$ is also convex and $\mathbf{x} \in \mathbb{R}^{d}$. Let $\mathbf{x}^{\star} \in \mathbb{R}^{d}$ be the minimizer of this function. Assume that $f_{t}$ are $L$-Lipschitz, where $L$ is constant, and that the distance between two points $\mathbf{x}$ and $\mathbf{x}'$ is bounded: $D(\mathbf{x}||\mathbf{x}') = \frac{1}{2}||\mathbf{x} - \mathbf{x}'||^{2}_{2} \leq F^{2}$, where $F$ is constant. This distance measure can be used to bound the magnitude of the differences between two states.

Let an update be given by $\mathbf{u}_{t} = -\eta_{t} \nabla f_{t}(\tilde{\mathbf{x}}_{t})$ and the  learning rate by $\eta_{t} = \frac{\sigma}{\sqrt{t}}$ with constant $\sigma$. Here, $\sigma$ can be used to adjust the learning rate to ensure convergence.

Represent the lag of updates due to network overheads and the different execution speeds of the $P$ workers by a vector, $\mathbf{\gamma}_{t} \in \mathbb{R}^{d}$, which consists of random variables, $Y_{i}$. These $Y_{i}$ are assumed i.i.d and independent of $\mathbf{u}_{t}$ and $\tilde{\mathbf{x}}_{t}$. Assume the same distribution for each $\gamma_{t}$ so that $\forall t .\mathbb{E}(\gamma_{t}) = \mu$ and $\forall t. \mathbb{E}(\gamma_{t}^{2}) = \phi$. That is, they have constant mean and variance. 

Following the presentation of regret by Ho et al.~\cite{ho2013}, let $R[X] =  \sum_{t}^{T} f_{t}(\tilde{\mathbf{x}}_{t}) - f_{t}(\mathbf{x}^{\star})$. This is the sum of the differences between the optimal value of the function and the current value given a noisy state. A probabilistic bound on the regret allows us to infer if the noisy system state, $\tilde{\mathbf{x}}_{t}$, is expected to converge towards the optimal, $\mathbf{x}^{\star}$, in probability. One such bound on the regret is given by:
\begin{equation}
\mathbb{P}\left ( \frac{R[X]}{T} - \frac{1}{\sqrt{T}} \left( \sigma L^{2} - \frac{2F^{2}}{\sigma} \right )- 4P\sigma L  \mu  \geq \delta \right ) \leq \exp \left ( -\frac{T\delta^{2}}{16P^{2}\sigma^{2} L^{2}\phi + \frac{b\delta}{3}} \right )\,,
\end{equation} for constant $\delta$ and $b \leq 4PT\sigma L$.

The $b$ term here is the upper bound on the random variables which are drawn from the lag distribution. If we assume with probability $\Phi$ that $\forall t. 4PL\sigma \gamma_{t} < O(T)$, then $b < O(T)$ so $\frac{R[X]}{T}$ converges to $O(T^{-1/2})$ in probability with an exponential tail bound with probability $\Phi$.

\subsection*{Proof}
The start of the proof proceeds as in Dai et al.~\cite{dai2014} and Ho et al.~\cite{ho2013}. After some manipulation, the regret term reduces into some deterministic terms and a probabilistic term. we find bounds for the deterministic terms and then the probabilistic term. Finally, we use a one-sided Bernstein inequality, and the bound on the regret, to show that ASP will converge in probability.

\paragraph{Manipulating the regret} Starting with the definition of the regret:
\begin{align}
R[X] &=  \sum_{t=1}^{T} f_{t}(\tilde{\mathbf{x}}_{t}) - f_{t}(\mathbf{x}^{\star}) \\
& \leq \sum_{t=1}^{T} \langle \nabla f_{t}(\tilde{\mathbf{x}}_{t}), \tilde{\mathbf{x}}_{t} - \mathbf{x}^{\star} \rangle  \text{     (by } f_{t} \text{ convex)} \\ \label{eq:aspregretsum}  &=  \sum_{t=1}^{T} \langle  \tilde{\mathbf{g}_{t}}, \tilde{\mathbf{x}}_{t} - \mathbf{x}^{\star} \rangle \,.
\end{align}
Here, $\tilde{\mathbf{g}_{t}} = \nabla f_{t}(\tilde{\mathbf{x}}_{t})$ and $\langle \mathbf{a}, \mathbf{b} \rangle$ is the dot product of $\mathbf{a}$ and $\mathbf{b}$.

Now, find a bound so that $R[X] < O(T)$, meaning that $\mathbb{E}(f_{t}(\tilde{\mathbf{x}}_{t}) - f_{t}(\mathbf{x}^{\star})) \rightarrow 0$ as $t \rightarrow \infty$. A bound involving steps $t$ and $t+1$ is particularly useful. Ho et al.~\cite{ho2013} prove the following lemma:

\textbf{Lemma 1}
Let $\tilde{\mathbf{g}}_{t}$ be as defined above, $D(\mathbf{x} || \mathbf{x}') = \frac{1}{2}||\mathbf{x} - \mathbf{x}'||^{2}_{2} $, and $\mathcal{A}_{t}$ and $\mathcal{B}_{t}$ are the index sets. The dot product between $\tilde{\mathbf{g}}_{t}$ and the difference between the current state, $\tilde{\mathbf{x}}_{t}$, and the optimal, $\mathbf{x}^{\star}$, is given by:
\begin{equation*}
\langle \tilde{\mathbf{g}}_{t}, \tilde{\mathbf{x}}_{t} - \mathbf{x}^{\star}\rangle = \frac{1}{2}\eta_{t} ||\tilde{\mathbf{g}_{t}}||^{2} + \frac{D(\mathbf{x}^{\star}|| \mathbf{x}_{t}) - D(\mathbf{x}^{\star} || \mathbf{x}_{t+1})}{\eta_{t}} + \left [ \sum_{i \in \mathcal{A}_{t}} \eta_{i}\langle \tilde{\mathbf{g}}_{i}, \tilde{\mathbf{g}}_{t} \rangle - \sum_{i \in \mathcal{B}} \eta_{i} \langle \tilde{\mathbf{g}}_{i}, \tilde{\mathbf{g}}_{t} \rangle  \right ]\,.
\end{equation*}
The first term incorporates the updates from the current iteration, the second is the distance between two successive states at $t$ and $t+1$ compared to the optimal $\mathbf{x}^{\star}$, and the final term incorporates the extra updates and missed updates at the current step.
\begin{flushright}
$\qedsymbol$
\end{flushright}

By application of Lemma 1:
\begin{align*}
R[X] \leq  \sum_{t=1}^{T}  \langle \tilde{\mathbf{g}}_{t}, \tilde{\mathbf{x}}_{t} - \mathbf{x}^{\star}\rangle =&  \sum_{t=1}^{T} \left [ \frac{1}{2}\eta_{t} ||\tilde{\mathbf{g}_{t}}||^{2} +  \frac{D(\mathbf{x}^{\star}|| \mathbf{x}_{t}) - D(\mathbf{x}^{\star} || \mathbf{x}_{t+1})}{\eta_{t}} \right ] \\
 & +  \sum_{t=1}^{T}  \left [ \sum_{i \in \mathcal{A}_{t}} \eta_{i}\langle \tilde{\mathbf{g}}_{i}, \tilde{\mathbf{g}}_{t} \rangle - \sum_{i \in \mathcal{B}} \eta_{i} \langle \tilde{\mathbf{g}}_{i}, \tilde{\mathbf{g}}_{t} \rangle  \right ]  \\ 
\label{eq:aspproofregretterm} =&  \sum_{t=1}^{T}  \left [  \frac{1}{2}\eta_{t} ||\tilde{\mathbf{g}_{t}}||^{2}  \right ] \\
&  + \left [ \frac{D(\mathbf{x}^{\star}|| \mathbf{x}_{1})}{\eta_{1}}- \frac{D(\mathbf{x}^{\star} || \mathbf{x}_{T+1})}{\eta_{T}}  +   \sum^{T}_{t=2} D(\mathbf{x}^{\star}||\mathbf{x}_{t}) \left ( \frac{1}{\eta_{t}} - \frac{1}{\eta_{t-1}} \right ) \right ] \\
& +  \sum_{t=1}^{T}  \left [ \sum_{i \in \mathcal{A}_{t}} \eta_{i}\langle \tilde{\mathbf{g}}_{i}, \tilde{\mathbf{g}}_{t} \rangle - \sum_{i \in \mathcal{B}_{t}} \eta_{i} \langle \tilde{\mathbf{g}}_{i}, \tilde{\mathbf{g}}_{t} \rangle   \right ]\,.
\end{align*}

\paragraph{Deterministic bounds}  we now find bounds for each term of $R[X]$.

Starting with the first term in the regret:
\begin{align*}
\sum_{t=1}^{T}\frac{1}{2}\eta_{t} ||\tilde{\mathbf{g}_{t}}||^{2} & \leq \sum_{t=1}^{T} \frac{1}{2}\eta_{t}L^{2} \textrm{ (by Lipschitz continuity) } \\
 & \leq \sum_{t=1}^{T}\frac{1}{2}\frac{\sigma}{\sqrt{t}}L^{2} \\
 & \leq \sigma L^{2} \sqrt{T}\,.
\end{align*}

The second term:
\begin{align*}
&\frac{D(\mathbf{x}^{\star}|| \mathbf{x}_{1})}{\eta_{1}}- \frac{D(\mathbf{x}^{\star} || \mathbf{x}_{T+1})}{\eta_{t}}  +   \sum^{T}_{t=2} D(\mathbf{x}^{\star}||\mathbf{x}_{t}) \left ( \frac{1}{\eta_{t}} - \frac{1}{\eta_{t-1}} \right ) \\ & \leq  \frac{F^{2}}{\sigma} + \frac{F^{2}\sqrt{T}}{\sigma} + \frac{F^{2}}{\sigma} \sum_{t=2}^{T} \left ( \sqrt{t} - \sqrt{t-1} \right ) \\
& \leq \frac{F^{2}}{\sigma} + \frac{F^{2}\sqrt{T}}{\sigma} + \frac{F^{2}}{\sigma} (\sqrt{T} - 1) \\
& \leq \frac{2F^{2}}{\sigma}\sqrt{T}\,.
\end{align*}
 
Now for the final term. We first provide a general form which is applicable to both ASP and PSP.  The final bounds require assumptions over the distribution of lags. Let $\bar{u}_{t} = \frac{1}{|\mathcal{A}||\mathcal{B}|} \sum_{i \in \mathcal{A} \cup \mathcal{B}} || \tilde{\mathbf{u}}_{t} ||_{2}$ be the average $l^{2}$-norm of the updates. The update equation is now (see Ho et al.~\cite{ho2013}):
\begin{equation}
\tilde{\mathbf{x}}_{t} = \mathbf{x}_{t} + \bar{u}_{t}\mathbf{\gamma}_{t}\,.
\end{equation}
 With this and $\bar{u}_{t}$, we can simplify the notation as follows:
\begin{align*}
 \sum_{t=1}^{T}  \left [  \sum_{i \in \mathcal{A}_{t}} \eta_{i}\langle \tilde{\mathbf{g}}_{i}, \tilde{\mathbf{g}}_{t} \rangle - \sum_{i \in \mathcal{B}_{t}} \eta_{i} \langle \tilde{\mathbf{g}}_{i}, \tilde{\mathbf{g}}_{t} \rangle    \right ] 
 \leq  & \sum_{t=1}^{T} \eta_{t} \langle \bar{u}_{t} \mathbf{\gamma}_{t}, \tilde{\mathbf{g}}_{t} \rangle \\
 \leq & \sum_{t=1}^{T} \eta_{t} \bar{u}_{t} ||\mathbf{\gamma}_{t}||_{2} ||\tilde{\mathbf{g}}_{t}||_{2}\,.\\
\end{align*}
This final term has many different components which require bounds. We now prove several lemmas for this purpose.

\textbf{Lemma 2 }
The average $l^{2}$-norm of the updates is bounded: $\bar{u}_{t} \leq 4PL \sqrt{t}$. Additionally, $|| \tilde{\mathbf{u}}_{t}||_{2} = || \eta_{t}\tilde{\mathbf{g}}_{t} || \leq  \eta_{t} L$ by Lipschitz continuity.  By definition:
\begin{align*}
\bar{u}_{t}  = &\frac{1}{|\mathcal{A}||\mathcal{B}|} \sum_{i \in \mathcal{A} \cup \mathcal{B}} || \tilde{\mathbf{u}}_{i} ||_{2}  \\
= & \frac{1}{|\mathcal{A}||\mathcal{B}|} \sum_{i \in \mathcal{A} \cup \mathcal{B}} || \eta_{i}\tilde{\mathbf{g}}_{i} ||_{2}  \\
\leq  & \frac{1}{|\mathcal{A}||\mathcal{B}|} \sum_{i \in \mathcal{A} \cup \mathcal{B}}  \eta_{i} L\,.
\intertext{
The worst case is if all updates are missed. Thus, bounds for the size of the index sets are given by: $1 \leq |\mathcal{A}|  \leq Pt$ and $1 \leq |\mathcal{B}| \leq Pt$ as $t$ steps have been performed and $P$ workers are present. Note that the total sizes of the two sets is going to be $\leq Pt$ but we ignore this in the following bound. So,
}
\bar{u}_{t} \leq & \frac{1}{|\mathcal{A}||\mathcal{B}|} 2PL \sum_{j=0}^{t} \frac{\sigma}{\sqrt{j}} \\
 \leq & 2PL \sum_{j=0}^{t}  \frac{\sigma}{\sqrt{j}}\,. \\
\intertext{Using the identity $ \sum_{i=a}^{b} \frac{1}{\sqrt{i}} \leq 2 \sqrt{b - a -1}$,}
\bar{u}_{t} \leq & 4PL \sqrt{t - 1}  \\ 
\leq & 4PL \sqrt{t}\,.
\end{align*}

\begin{flushright}
$\qedsymbol$
\end{flushright}

\textbf{Lemma 3 } The L2 norm of $\gamma_{t}$ is bounded: $||\gamma_{t}||_{2} \leq TP $. 

In the worst case, an update can experience a lag of $T$ and there are $P$ workers who could lag. If all $P$ entries of $\gamma_{t}$ are $T$ then $||\gamma_{t}||_{2} \leq TP$.

\begin{flushright}
$\qedsymbol$
\end{flushright}

Back to the theorem. Using Lemma 2 on the final term of the regret yields:
\begin{align*}
\sum_{t=1}^{T} \eta_{t} \bar{u}_{t} ||\gamma_{t}||_{2} ||\tilde{g}_{t}||_{2}  \leq & \sum_{t=1}^{T} \frac{\sigma}{\sqrt{t}} 4PL \sqrt{t} L ||\gamma_{t}||^{2} \\
 = & 4PL \sigma \sum_{t=1}^{T} \gamma_{t}\,.
 \end{align*}

Substituting the new bounds back into $R[X]$ yields:
\begin{align*}
R[X] \leq \sigma L^{2} \sqrt{T} + \frac{2F^{2}}{\sigma}\sqrt{T} + 4P\sigma L \sum_{t=1}^{T} \gamma_{t}\,.
\end{align*}

Dividing through by $T$:
\begin{equation} \label{eq:psprnearbound}
\frac{R[X]}{T} - \frac{\sigma L^{2}}{\sqrt{T}} - \frac{2F^{2}}{\sigma \sqrt{T}} \leq \frac{4P\sigma L\sum_{t=1}^{T} \gamma_{t}}{T}\,.
\end{equation} 

This represents a bound on the regret.  However, the bound from Lemma 3 on $\gamma_{t}$ would lead to an upper bound of $4PL\sigma \sum_{t=1}^{T} TP$ which is $O(T^{2})$. Clearly, this leads to $R[X] = O(T^{2})$ which means we cannot guarantee convergence. However, if stronger assumptions are made on the distribution of $\gamma_{t}$, probabilistic bounds can be found.

\paragraph{Probabilistic Bound}
 To bound the regret in a more conservative fashion than Lemma 3, it is necessary to consider a probabilistic bound. Denote $Z_{t}$ as a random variable where each $Z_{t}$ is bounded, $Z_{t} \leq b \leq 4PL\sigma T$, and independent. A one-sided Bernstein inequality  from Pollard~\cite{pollard1984} is:
\begin{align*}
\mathbb{P}\left ( \sum_{t=1}^{T} (Z_{t} - \mathbb{E}(Z_{t})) \geq T\delta \right ) \leq \exp \left ( -\frac{T\delta^{2}}{\frac{1}{T}\sum_{t=1}^{T} \mathbb{E}(Z_{t}^{2}) + \frac{b\delta}{3}} \right )\,,
\end{align*} where $\delta$ is a constant. This can be used to bound the differences between the mean of the distribution and the random variables which will enable the derivation of a bound on the regret.

Letting $Z_{t} = 4P\sigma L \gamma_{t}$ yields:
 \begin{align*}
\mathbb{P}\left ( \frac{\sum_{t=1}^{T} 4P\sigma L\gamma_{t} -  \sum_{t=1}^{T}  \mathbb{E}(4P\sigma L\gamma_{t})}{T} \geq \delta \right ) \leq \exp \left ( -\frac{T\delta^{2}}{\frac{1}{T}\sum_{t=1}^{T} \mathbb{E}((4P\sigma L\gamma_{t})^{2})  + \frac{b\delta}{3}} \right )\,.
\end{align*}

Using \ref{eq:psprnearbound}, the regret can be introduced into the expression:
 \begin{align*}	
\mathbb{P}\left ( \frac{R[X]}{T} - \frac{1}{\sqrt{T}} \left( \sigma L^{2} - \frac{2F^{2}}{\sigma} \right ) - \frac{4P\sigma L}{T}  \left  ( \sum_{t=1}^{T}  \mathbb{E}(\gamma_{t}) \right ) \geq \delta \right ) \leq & \exp \left ( -\frac{T\delta^{2}}{\frac{16P^{2}\sigma^{2} L^{2}}{T} \sum_{t=1}^{T}  \mathbb{E}(\gamma_{t}^{2})  + \frac{b\delta}{3}} \right )\,.
\end{align*}
\textit{Convergence is dependent upon the mean and variance of the lag distribution.
However, for distributions where $\sum_{t=1}^{T}\mathbb{E}(\gamma_{t}) \leq \sqrt{T}$, $\frac{R[X]}{T}$ converges to $O(T^{-1/2})$ with an exponential tail bound.} This implies probabilistic convergence. Unfortunately, this limit on the sum of the means is rather tight. Instead, let us assume the same distribution for each $\gamma_{t}$ so that $\forall t .\mathbb{E}(\gamma_{t}) = \mu$ and $\forall t. \mathbb{E}(\gamma_{t}^{2}) = \phi$, that is constant mean and variance. We can relax our bound by letting  $\delta' = \delta + 4PL\mu\sigma$, which is constant:
\begin{align*}
\mathbb{P}\left ( \frac{R[X]}{T} - \frac{1}{\sqrt{T}} \left( \sigma L^{2} - \frac{2F^{2}}{\sigma} \right ) - 4P\sigma L  \mu \geq \delta \right ) \leq & \exp \left ( -\frac{T\delta^{2}}{16P^{2}\sigma^{2} L^{2}\phi + \frac{b\delta}{3}}  \right )  \\\\
\mathbb{P}\left ( \frac{R[X]}{T} - \frac{1}{\sqrt{T}} \left( \sigma L^{2} - \frac{2F^{2}}{\sigma} \right ) \geq \delta' \right ) \leq & \exp \left ( -\frac{T\delta'^{2}}{16P^{2}\sigma^{2} L^{2}\phi  + \frac{b\delta}{3}} \right )\,. \\ 
\end{align*}
If we assume with probability $\Phi$ that $\forall t. 4PL\sigma \gamma_{t} < O(T)$, then $b < O(T)$ so $\frac{R[X]}{T}$ converges to $O(T^{-1/2})$ in probability with an exponential tail bound with probability $\Phi$.

\begin{flushright}
$\qedsymbol$
\end{flushright}

\subsection{Convergence of SGD under PSP}

In PSP, either a central oracle tracks the progress of each worker or the workers each hold their own local view. At the barrier control point, a worker samples $\beta$ out of $P$ workers without replacement. If a single one of these lags more than $s$  steps behind the current worker then it waits. This process is pBSP (based on BSP) if $s = 0$ and pSSP (based on SSP) if $s > 0$. However, if $s = \infty$ then PSP reduces to ASP.

\textit{PSP improves on ASP by providing probabilistic guarantees about convergence with tighter bounds and less restrictive assumptions. pSSP relaxes SSP's inflexible staleness parameter by permitting some workers to fall further behind. This works because machine learning algorithms can typically tolerate the additional noise~\cite{cipar2013}.  pBSP relaxes BSP by allowing some workers to lag slightly,  yielding a BSP-like method which is more resistant to stragglers but no longer deterministic.} We now formalise the PSP method.

\subsection*{Theorem 2: PSP sampling}
Take a set of $P$ workers whose probabilities of lagging $r$ steps are drawn from a distribution with probability mass function $f(r)$ and cumulative distribution function (CDF) $F(r) = \mathbb{P}(x \leq r) = \sum_{x=0}^{r} f(r)$.  Sample without replacement $\beta$ workers where $\beta \in \mathbb{Z}^{+}$ and $\beta \leq P$. Impose the sampling constraint that if a single one of the $\beta$ workers has lag greater than r then the sampler must wait. This leads to the following distribution for lags:
\begin{align*}
p(s) = 
\left\{\begin{array}{lr}
\alpha f(s) & \text{ for } s \leq r  \\
\alpha(F(r)^{\beta})^{s-r} & \text{ for } s > r
 \end{array}\right\}\,,
\end{align*}\\
where $\alpha$ is some normalising constant.
\subsection*{Proof}
Let $n$ be one of the sampled workers. For $s \leq r$, the probability that a worker will lag $s$ steps depends entirely upon its lag distribution as the \sampling primitive
only considers workers with $s > r$. 

For $s > r$, a worker waits if at least one of the $\beta$ workers lags more than $r$ steps and otherwise it proceeds. The probability that a worker's step count increases by one beyond $r$ is thus given by:
\begin{align*}
\mathbb{P}(\forall n. \text{lag}(n) \leq r) = & F(r)^{\beta}\,.
\end{align*}
This assumes that the probability of a worker lagging is independent of the state of the other workers. In order for a worker to lag $s-r$ steps, where $s > r$, it has to have been missed $s-r$ times by the sampling primitive. Assuming each sampling event is independent yields:
\begin{align*}
\forall n. \forall s > r. \mathbb{P}(\text{lag}(n) = s) = (F(r)^{\beta})^{s-r}\,.
\end{align*}
Now, to make $p(s)$ a valid probability distribution, it needs to be normalised over $[0, T]$ inclusive:
\begin{equation} \label{eq:psptheoryalpha}
\alpha \sum_{s=0}^{\infty} p(s) = 1\,.
\end{equation}
Evaluating the sum:
\begin{align}
\sum_{s=0}^{T} p(s) = & \sum_{s=0}^{r} f(s) + \sum_{s=r+1}^{T}(F(r)^{\beta})^{s-r}  \\
\label{eq:pspsumtheory} = &  \sum_{s=0}^{r} f(s)  + \sum_{s=1}^{T-r}(F(r)^{\beta})^{s}\,.  \\
\intertext{If $F(r)^{\beta} < 1$ then we have a geometric series:}
\sum_{s=0}^{T} p(s) = & \sum_{s=0}^{r} f(s)  + \frac{(1 -(F(r)^{\beta})^{T-r+1})}{1-F(r)^{\beta}} -1\,.
\intertext{Otherwise, if $F(r)^{\beta} = 1$:}
\sum_{s=0}^{T} p(s) = &  \sum_{s=0}^{r} f(s)+  T -r\,.
\end{align}
A substitution and rearrangement yields $\alpha$ in \ref{eq:psptheoryalpha}.

To bound $\alpha$, examine the sum in \ref{eq:pspsumtheory}. For $T > r + 1$, 
\begin{equation}
\sum_{s=0}^{T} p(s) \leq F(r) + F(r)^{\beta}\,,
\end{equation}
because each term of the geometric sum is positive.  Thus, by \ref{eq:psptheoryalpha}:
\begin{equation}
\alpha \geq \frac{1}{F(r) + F(r)^{\beta}}\,.
\end{equation}
If $F(r)^{\beta} = 1$ then,
\begin{equation}
\alpha \leq \frac{1}{T-r}\,.
\end{equation}
\begin{flushright}
$\qedsymbol$
\end{flushright}

\subsection*{Theorem 3: SGD under PSP}
Let $f(\mathbf{x}) = \sum_{t=1}^{T} f_{t}(\mathbf{x})$ be a convex function where each $f_{t} \in \mathbb{R}$ is also convex. Let $\mathbf{x}^{\star} \in \mathbb{R}^{d}$ be the minimizer of this function. Assume that $f_{t}$ are L-Lipschitz and that the distance between two points $\mathbf{x}$ and $\mathbf{x}'$ is bounded: $D(\mathbf{x}||\mathbf{x}') = \frac{1}{2}||\mathbf{x} - \mathbf{x}'||^{2}_{2} \leq F^{2}$, where $F$ is constant. 

Let an update be given by $\mathbf{u}_{t} = -\eta_{t} \nabla f_{t}(\tilde{\mathbf{x}}_{t})$ and the learning rate by $\eta_{t} = \frac{\sigma}{\sqrt{t}}$.

Represent the lag of updates due to network overheads and the different execution speeds of the $P$ workers by a vector, $\mathbf{\gamma}_{t} \in \mathbb{R}^{d}$, which consists of random variables, $Y_{i}$. These $Y_{i}$ are i.i.d and are independent of $\mathbf{u}_{t}$ and $\tilde{\mathbf{x}}$. 
	
Following the presentation of regret by Ho et al.~\cite{ho2013}, 	let $R[X] =  \sum_{t}^{T} f_{t}(\tilde{\mathbf{x}}_{t}) - f_{t}(\mathbf{x}^{\star})$.  This is the sum of the differences between the optimal value of the function and the current value given a noisy state. A probabilistic bound on the regret allows us to determine if the noisy system state, $\tilde{\mathbf{x}}_{t}$, converges towards the optimal, $\mathbf{x}^{\star}$, in probability. One such bound on the regret is given by:
 \begin{equation}\label{eq:psptheorem}
\mathbb{P}\left ( \frac{R[X]}{T} - \frac{1}{\sqrt{T}} \left( \sigma L^{2} - \frac{2F^{2}}{\sigma} \right ) - q \geq \delta \right ) \leq \exp \left ( -\frac{T\delta^{2}}{c  + \frac{b\delta}{3}} \right )\,,
\end{equation} where $\delta$ is a constant and $b \leq 4PTL \sigma$. The $b$ term here is the upper bound on the random variables which are drawn from the lag distribution. 

Let $a = F(r)^{\beta} =  (\sum_{s=0}^{r} p(s))^{\beta}$. If we assume that $0 < a < 1$, then:
\begin{align}
q& \leq  \frac{4P\sigma L (1-a)}{F(r)(1-a) + a - a^{T-r+1}} \left ( \frac{r(r+1)}{2} + \frac{  a(r + 2)}{(1-a)^{2}} \right)\,.
\end{align}
Again, assuming that $0 < a < 1$, then the value of $c$ is bounded by the following expression:
\begin{align}
c  \leq &  \frac{16P^{2}\sigma^{2} L^{2}(1-a)}{F(r)(1-a) + a - a^{T-r+1}} \biggl (\frac{r(r+1)(2r+1)}{6} + \frac{a(r^{2} + 4)}{(1-a)^{3}}\biggr )\,.
\end{align}

If we further assume with probability $\Phi$ that $\forall t. 4PL\sigma \gamma_{t} < O(T)$, then $b < O(T)$ so $\frac{R[X]}{T}$ converges to $O(T^{-1/2})$ in probability with an exponential tail bound with probability $\Phi$.

\subsection*{Proof}
The initial parts of our convergence proof for ASP provide the starting point for our analysis of PSP. We begin with equation \ref{eq:psprnearbound} in Theorem 1 and construct one-sided probabilistic Bernstein bounds on the regret. Next, we consider how PSP impacts this bound by examining how it bounds the average of the means and then the average of the variances.

\paragraph{Probabilistic Bound}
Following the application of Lemma 2 in Theorem 1, we have:
\begin{equation} 
\frac{R[X]}{T} - \frac{\sigma L^{2}}{\sqrt{T}} - \frac{2F^{2}}{\sigma \sqrt{T}} \leq \frac{4P\sigma L\sum_{t=1}^{T} \mathbf{\gamma}_{t}}{T}\,.
\end{equation} 
 
Assume $Z_{t} \leq b \leq 4PTL \sigma $ and that each $Z_{t}$ is independent. Then the following one-sided Bernstein inequality from Pollard~\cite{pollard1984}  can be used:
\begin{equation}
\mathbb{P}\left ( \sum_{t=0}^{T} (Z_{t} - \mathbb{E}(Z_{t})) \geq T\delta \right ) \leq \exp \left ( -\frac{T\delta^{2}}{\frac{1}{T}\sum_{t=0}^{T} \mathbb{E}(Z_{t}^{2}) + \frac{b\delta}{3}} \right )\,.
\end{equation}
Using the bounds in Lemma 3, let $Z_{t} = 4P\sigma L \mathbf{\gamma}_{t}$ so,
 \begin{align*}
\mathbb{P}\left ( \frac{\sum_{t=0}^{T} 4P\sigma L\gamma_{t} -  \sum_{t=0}^{T}  \mathbb{E}(4P\sigma L\gamma_{t})}{T} \geq \delta \right ) \leq \exp \left ( -\frac{T\delta^{2}}{\frac{1}{T}\sum_{t=0}^{T} \mathbb{E}((4P\sigma L\gamma_{t})^{2}) + \frac{b\delta}{3}} \right )\,.
\end{align*}
Substituting in equation \ref{eq:psprnearbound},
 \begin{align*}
\mathbb{P}\left ( \frac{R[X]}{T} - \frac{1}{\sqrt{T}} \left( \sigma L^{2} - \frac{2F^{2}}{\sigma} \right ) - \frac{4P\sigma L}{T}  \left  ( \sum_{t=1}^{T}  \mathbb{E}(\gamma_{t}) \right ) \geq \delta \right ) \leq & \exp \left ( -\frac{T\delta^{2}}{\frac{16P^{2}\sigma^{2} L^{2}}{T}\sum_{t=0}^{T} \mathbb{E}(\gamma_{t}^{2})  + \frac{b\delta}{3}} \right )\,.
\end{align*}
Clearly the probability is dependent upon the mean, $ \mathbb{E}(\mathbf{\gamma}_{t})$, and variance, $ \mathbb{E}(\mathbf{\gamma}_{t}^{2})$, of the lag distribution. Currently, we need $\sum_{t=1}^{T}\mathbb{E}(\gamma_{t}) \leq \sqrt{T}$. However, this limit on the sum of the means is unfortunately tight. In ASP, an assumption that the mean and variance were constant yielded a bound but we had to increase our constant, $\delta$, by adding $4PL\mu\sigma$ which could make the constant far too large to be practical. For example, if $\mu$ is $T$. The aim is to see how using the PSP \sampling primitive impacts this bound. We focus first on the term with the means and then the term with the variances.

\paragraph{Bounding the average of the means} 
First the $\frac{1}{T} \sum_{t=1}^{T}  \mathbb{E}(\mathbf{\gamma}_{t})$ term:
\begin{align}
 \frac{1}{T}  \sum_{t=0}^{T}  \mathbb{E}(\mathbf{\gamma}_{t}) = & \frac{1}{T} \sum_{t=0}^{T} \sum_{s=1}^{t}p(s)s\,.
 \end{align}
 
By the definition of PSP in Theorem 2,
\begin{align}
\label{eq:pspmeansum}  \frac{1}{T}  \sum_{t=0}^{T}  \mathbb{E}(\mathbf{\gamma}_{t}) =  & \frac{\alpha}{T} \left ( \sum_{t=0}^{r} \left ( \sum_{s=0}^{t}f(s)s \right ) +  \sum_{t=r+1}^{T}\left ( \sum_{s=0}^{r}f(s)s + \sum_{s = r+1}^{t}(F(r)^{\beta})^{s-r}s \right ) \right ) \,.
\end{align}

Letting $a = F(r)^{\beta} = \left (\sum_{i=1}^{r}p(i) \right )^{\beta}$ where $0 \leq a \leq 1$, bounding summations, and performing some rearrangements:
\begin{align}
\frac{1}{T}  \sum_{t=0}^{T}  \mathbb{E}(\gamma_{t})
\leq &  \frac{\alpha}{T} \left ( r \sum_{s=0}^{r}f(s)s +  \sum_{t=r+1}^{T}\left ( \sum_{s=0}^{r}f(s)s + \sum_{s = r+1}^{t} a^{s-r}s \right ) \right ) \\
\leq & \frac{\alpha}{T} \left ( T \sum_{s=0}^{r}f(s)s +  \sum_{t=r+1}^{T}\left (\sum_{s = r+1}^{t} a^{s-r}s \right ) \right ) \\
\leq & \alpha \sum_{s=0}^{r}f(s)s + \frac{\alpha}{T}   \sum_{t=r+1}^{T}\left (\sum_{s = r+1}^{t} a^{s-r}s \right )\,.\\
\intertext{The indexing of the inner sum over $[r+1,t]$ can be rewritten so that a closed-form solution can be used:}
\label{eq:pspprearithm} \leq & \alpha \sum_{s=0}^{r}f(s)s + \frac{\alpha}{T}   \sum_{t=r+1}^{T}\left (a\sum_{s = 1}^{t-r} a^{s-1} (s+r) \right )\,.
\end{align}
Specifically, the inner summation on $[1,t-r]$ is over an arithmetico-geometric series. 

\subparagraph{Arithmetico-geometric series} An arithmetico-geometric series takes the following form:
\begin{equation}\label{eq:arithmeticogeodef}
S_{n} = \sum_{k=1}^{n}(a + (k-1)d)r^{k-1}\,.
\end{equation}

There exists a closed-form solution to this partial sum. See D. Khattar's~\cite{khattar} book for the proof. 

Provided $r \neq 1$ then,
\begin{equation} \label{eq:arithmeticogeo}
S_{n} = \frac{a - (a+(n-1)d)r^{n}}{1-r} + \frac{dr(1-r^{n-1})}{(1-r)^{2}}\,.
\end{equation} 

If $r = 1$ then,
\begin{equation} \label{eq:arithmeticogeoreq1}
S_{n} = \frac{n}{2}(2a + (n-1)d)\,.
\end{equation}
\begin{flushright}
$\qedsymbol$
\end{flushright}

Back to the proof. Make the substitutions $a = r+1$, $d = 1$, $n = t-r$ and $r = a$ in equation \ref{eq:arithmeticogeodef}.

For the first case in the arithmetico-geometric solution we have $r \neq 1$. Thus, $a < 1$ and $(1- a^{t-r-1}) \leq 1$. Substituting this result into the bound yields:
\begin{align}
\frac{1}{T}  \sum_{t=0}^{T}  \mathbb{E}(\gamma_{t})
\leq & \alpha \sum_{s=0}^{r}f(s)s + \frac{\alpha a}{T}  \sum_{t=r+1}^{T}\left ( \frac{r+1 - ta^{t-r}}{1-a} + \frac{a(1-a^{t-r-1})}{(1-a)^{2}} \right ) \\
\label{eq:pspproofpretrsub} \leq & \alpha \sum_{s=0}^{r}f(s)s + \frac{\alpha a}{T}  \left ( \frac{a(T-r)}{(1-a)^{2}} + \frac{(T-r)(r+1)}{1-a} - \sum_{t=r+1}^{T}\left ( \frac{ ta^{t-r}}{1-a} \right ) \right )\,.
\end{align}
Examining $ \sum_{t=r+1}^{T}\left ( \frac{ ta^{t-r}}{1-a} \right ) $. For $T > r+1$, this summation is at least $\frac{Ta^{T-r}}{1-a}$ as each term in the summation is positive. Removing some negative terms and using the bound on the sum:
\begin{align}
\frac{1}{T}  \sum_{t=0}^{T}  \mathbb{E}(\gamma_{t}) & \leq \alpha \sum_{s=0}^{r}f(s)s + \frac{\alpha a }{T} \left  ( \frac{T(a + (1-a)(r+1) - (1-a)a^{T-r})}{(1-a)^{2}}  \right ) 
\end{align}
\begin{align}
\frac{1}{T}  \sum_{t=0}^{T}  \mathbb{E}(\gamma_{t}) & \leq \alpha \sum_{s=0}^{r}f(s)s + \frac{\alpha  a}{(1-a)^{2}} \left ( a+(1-a)(r+1) - (1-a)a^{T-r}  \right ) \\
& \label{eq:pspproofbeforealpha} \leq \alpha \sum_{s=0}^{r}f(s)s + \frac{\alpha  a}{(1-a)^{2}} \left ( r + 2 \right )\,.
\end{align}

By Theorem 2, $\alpha$ is defined as:
\begin{align}
\alpha & =  \frac{1-F(r)^{\beta}}{ F(r)( 1- F(r)^{\beta}) + (1- (F(r)^{\beta})^{T-r+1}) - (1 - F(r)^{\beta})} \\
 & = \frac{1-a}{F(r)(1-a) + a - a^{T-r+1}}\,.
\end{align}
The first term in \ref{eq:pspproofbeforealpha} can be bounded as $f(s) \leq 1$. Specifically, letting $\forall s. f(s) =1$ yields an arithmetic series. Taking the partial sum and substituting in $\alpha$:
\begin{align} \label{eq:pspmeanbound}
\frac{1}{T}  \sum_{t=0}^{T}  \mathbb{E}(\gamma_{t}) & \leq  \frac{1-a}{F(r)(1-a) + a - a^{T-r+1}} \left ( \frac{r(r+1)}{2} + \frac{  a(r + 2)}{(1-a)^{2}} \right)\,.
\end{align}

This is a bound on the average of the means of the lags which relies on the sampling count, $\beta$ (in the $a$ term), the staleness parameter, $r$, and the length of the update sequence, $T$.




Considering now the second case of the arithmetico-geometric series closed-form solution. Specifically with $r=1$ so that $a = 1$. Using \ref{eq:pspprearithm} we arrive at the following:
\begin{align}
\frac{1}{T}  \sum_{t=0}^{T}  \mathbb{E}(\gamma_{t})
\leq & \alpha\frac{r(r+1)}{2} + \frac{\alpha}{T}\sum_{t=r+1}^{T}\left ( \frac{t-r}{2}(2(r+1) + (t-r -1)) \right ) \\
\leq &  \alpha\frac{r(r+1)}{2} +   \frac{\alpha}{T}\left ( \frac{-1}{2}(T-r)(r^{2} + r) + \frac{1}{2}\sum_{t= r+1}^{T}  t^{2} + \frac{1}{2}\sum_{t=r+1}^{T}t \right )\,. \\
\intertext{Assuming $T > r$, removing some negative terms and substituting in solutions to the partial sums over squared arithmetic and arithmetic series yields:}
\frac{1}{T}  \sum_{t=0}^{T}  \mathbb{E}(\gamma_{t})\leq &  \alpha \frac{r(r+1)}{2} +   \frac{\alpha}{T} \biggl (\frac{T(T+1)(2T+1)}{12} \\ & + \frac{T(T+1)}{4} - \frac{r(r+1)(2r + 1)}{12} - \frac{r(r+1)}{4} \biggr )\,. \\
\intertext{By Theorem 2, as $a = 1$, $\alpha \leq \frac{1}{T-r}$ so, }
\frac{1}{T}  \sum_{t=0}^{T}  \mathbb{E}(\gamma_{t}) < &  \frac{1}{T-r} \biggl (  \frac{r(r+1)}{4} + \frac{(T+1)(2T+1)}{12} + \frac{T+1}{4} \biggr ) \\
< & \frac{1}{T-r} \biggl ( \frac{r(r+1)}{2} + T^{2}+T+Tr+r \biggr )\,.
\end{align}
This is $O(T)$, indicating that when $a = (\sum_{s=0}^{r} p(s))^{\beta} = 1$, PSP does not converge in probability. Thus, at least one worker needs to be sampled and some probability mass needs to be present in the first $r$ steps.

\paragraph{Bounding the average of the variances}
Now for the  $\frac{1}{T}\sum_{t=0}^{T}\mathbb{E}(\gamma_{t}^{2})$ term. This is the same process used to bound the average of the means but with $s^{2}$ rather than $s$ as the variable:
\begin{align}
 \frac{1}{T}  \sum_{t=0}^{T}  \mathbb{E}(\mathbf{\gamma}_{t}^{2}) 
\leq & \label{eq:pspvarprearithm} \alpha \sum_{s=0}^{r}f(s)s^{2} + \frac{\alpha}{T}   \sum_{t=r+1}^{T}\left (a\sum_{s = 1}^{t-r} a^{s-1} (s+r)^{2} \right )\,.
\end{align}
The inner summation on $[1,t-r]$ is no longer over an arithmetico-geometric series but a squared arithmetico-geometric series. 

\subparagraph{Squared Arithmetico-geometric series} The proof of the closed-form solution of the partial sum is skipped in this paper. This is given by:
\begin{equation} \label{eq:sqarithgeo}
S_{n} = 
\begin{cases}
\begin{aligned}[b]
& \biggl [ \frac{a^{2} - 2ad - d^{2} - (a + (n-1)d)^{2}r^{n}}{1-r}+ \frac{1-r^{n-1}}{(1-r)^{2}}\\
& + \frac{d^{2}}{1-r}\biggl ( \frac{1 - (1+ 2(n-2))r^{n-1}}{1-r} + \frac{2r(1 - r^{n-2})}{(1-r)^{2}}\biggr ) \biggr ],
\end{aligned} & \text{ for } |r| < 1
 \\
\begin{aligned}
na^{2} + \frac{2dn(n-1)}{2} + \frac{d^{2}n(n-1)(2n-1)}{6} \end{aligned}& \text { for } r = 1
\end{cases}
\end{equation}

For the case of $|r| < 1$ in equation \ref{eq:sqarithgeo}, we have $a < 1$. Make the substitutions: $a = r+ 1$, $d=1$, $n = t-r$ and $r = a$. Substituting this into \ref{eq:pspvarprearithm} yields:
\begin{align*}
 \frac{1}{T}  \sum_{t=0}^{T}  \mathbb{E}(\mathbf{\gamma}_{t}^{2}) 
\leq &  \alpha \sum_{s=0}^{r}f(s)s^{2} + \frac{\alpha a}{T}   \sum_{t=r+1}^{T} \biggl [ \frac{(r+1)^{2} - 2(r+1) - 1 - (r +1 + (t-r-1))^{2}a^{t-r}}{1-a} \\ & +  \frac{1-a^{t-r-1}}{(1-a)^{2}}\\
& + \frac{1}{1-a}\biggl ( \frac{1 - (1+ 2(t-r-2))a^{t-r-1}}{1-a} + \frac{2a(1 - a^{t-r-2})}{(1-a)^{2}}\biggr ) \biggr ] \\
\leq &  \alpha \sum_{s=0}^{r}f(s)s^{2} + \frac{\alpha a}{T}   \sum_{t=r+1}^{T} \biggl [ \frac{r^{2} -2 - t^{2}a^{t-r}}{1-a} +  \frac{1-a^{t-r-1}}{(1-a)^{2}}\\
& + \frac{1}{1-a}\biggl ( \frac{1 - (2t -2r -3)a^{t-r-1}}{1-a} + \frac{2a(1 - a^{t-r-2})}{(1-a)^{2}}\biggr ) \biggr ] \\
\leq &  \alpha \sum_{s=0}^{r}f(s)s^{2} + \frac{\alpha a}{T(1-a)^{3}}  \sum_{t=r+1}^{T}\biggl [  (1-a)^{2}(r^{2} - 2 - t^{2}a^{t-r}) + (1-a)(1-a^{t-r-2}) \\ & + (1-a)(1-(2t-2r-3)a^{t-r-1}) + 2a(1- a^{t-r-2}) \biggr ]\,. \\
\end{align*}
Assume $T > r+ 2$ and use $a < 1$ to bound some terms:
\begin{align*}
 \frac{1}{T}  \sum_{t=0}^{T}  \mathbb{E}(\mathbf{\gamma}_{t}^{2}) 
\leq &  \alpha \sum_{s=0}^{r}f(s)s^{2} + \frac{\alpha}{T}   \sum_{t=r+1}^{T} \biggl (\frac{a}{(1-a)^{3}}\biggl [ (1-a)^{2}r^{2} + (1-a) + (1-a) + 2a \biggr ] \biggr ) \\
\leq &  \alpha \sum_{s=0}^{r}f(s)s^{2} + \frac{\alpha a(T-r)}{T(1-a)^{3}}\biggl [ r^{2} + 4 \biggr ]\,.
\end{align*}

As $a < 1$, $\forall s. f(s) \leq 1$ so the first term can be bounded by setting $f(s) = 1$, yielding a squared arithmetic series. Take the partial sum:
\begin{align}
 \frac{1}{T}  \sum_{t=0}^{T}  \mathbb{E}(\mathbf{\gamma}_{t}^{2}) 
 < &  \alpha  \frac{r(r+1)(2r+1)}{6} + \frac{\alpha a(r^{2} + 4)}{(1-a)^{3}}\,.
\end{align}
Substituting in $\alpha$ from Theorem 2 yields a bound on the average of the variances of the lag distribution given by:
\begin{align} \label{eq:pspvariancebound}
\frac{1}{T}  \sum_{t=0}^{T}  \mathbb{E}(\mathbf{\gamma}_{t}^{2})   < &  \frac{1-a}{F(r)(1-a) + a - a^{T-r+1}} \biggl (\frac{r(r+1)(2r+1)}{6} + \frac{a(r^{2} + 4)}{(1-a)^{3}}\biggr )\,.
\end{align}

Consider now the case in \ref{eq:sqarithgeo} where $r=1$, meaning that $a=1$.  Make the substitutions: $a = r+ 1$, $d=1$, $n = t-r$ and $r = a$. This achieves:
\begin{align*}
\frac{1}{T}  \sum_{t=0}^{T}  \mathbb{E}(\mathbf{\gamma}_{t}^{2}) 
\leq &  \alpha \sum_{s=0}^{r}f(s)s^{2}  \\
& + \frac{\alpha a}{T}   \sum_{t=r+1}^{T} \biggl [  na^{2} + \frac{2dn(n-1)}{2} + \frac{d^{2}n(n-1)(2n-1)}{6} \biggr ]\\
\leq &  \alpha \sum_{s=0}^{r}f(s)s^{2} \\
& + \frac{\alpha a}{T}   \sum_{t=r+1}^{T} (t-r) \biggl [ (r+1)^{2} + \frac{2(t-r-1)}{2} + \frac{(t-r-1)(2(t-r)-1)}{6} \biggr ]\\
< &  \alpha \sum_{s=0}^{r}f(s)s^{2} + \frac{\alpha a}{T}   \sum_{t=r+1}^{T} \biggl [ t(r+1)^{2} + t^{2} + \frac{t^{3}}{3} \biggr ]\\
< &  \alpha \sum_{s=0}^{r}f(s)s^{2} + \alpha a \biggl [ \frac{(T+1)(r+1)^{2}}{2} + \frac{(T+1)(2T+1)}{6} + \frac{T(T+1)^{2}}{12} \biggr ]\,.
\end{align*}
Using $\alpha$ from Theorem 2, when $a =1$, $\alpha \leq \frac{1}{T-r}$, which provides a bound on the average of the variances of the lag distribution given by:
\begin{align*}
\frac{1}{T}  \sum_{t=0}^{T}  \mathbb{E}(\mathbf{\gamma}_{t}^{2})   < &  \frac{1}{T-r} \biggl ( \frac{r(r+1)(2r+1)}{6} \\ &+ a \biggl [\frac{(T+1)(r+1)^{2}}{2} + \frac{(T+1)(2T+1)}{6} + \frac{T(T+1)^{2}}{12}\biggr ] \biggr )\,.
\end{align*}
This is $O(T^{2})$ which indicates that if $a=1$, then SGD under PSP will not converge in probability. This is expected as $a=1$ if $\beta = 0$.

\begin{flushright}
$\qedsymbol$
\end{flushright}

\section{Discussion}

The following discussion extends our theoretical analysis and evaluation results to their real-world implications, and explains why PSP is a superior barrier control method to other existing solutions.

\subsection{Discussion of PSP Bounds} \label{sec:pspboundsdisc}
To conclude, we have shown that using PSP with SGD yields a probabilistic bound on the regret, implying convergence in probability. This happens in two scenarios. The first is when the average of the means and the average of the variances are constant, or bounded by a constant, causing the constant term in the probabilistic bound to be increased. The second case is that both sums are $ < O(\sqrt{T})$. 

We derived bounds for both the average of the means and the average of the variances. These can be treated as constants for fixed $a$, $T$, $r$ and $\beta$, meeting the first case for convergence in probability.

More specifically, the average of the means is bounded by:
\begin{align} \label{eq:pspmeanbounddis}
\frac{1}{T}  \sum_{t=0}^{T}  \mathbb{E}(\mathbf{\gamma}_{t}) & \leq  \frac{1-a}{F(r)(1-a) + a - a^{T-r+1}} \left ( \frac{r(r+1)}{2} + \frac{  a(r + 2)}{(1-a)^{2}} \right).
\end{align}
The average of the variances has a similar bound:
\begin{align} \label{eq:pspvariancebounddis}
\frac{1}{T}  \sum_{t=0}^{T}  \mathbb{E}(\mathbf{\gamma}_{t}^{2})   < &  \frac{1-a}{F(r)(1-a) + a - a^{T-r+1}} \biggl (\frac{r(r+1)(2r+1)}{6} + \frac{a(r^{2} + 4)}{(1-a)^{3}}\biggr ).
\end{align}

These bounds rely on the sampling count, $\beta$ (in $a$), the staleness parameter, $r$, the length of the update sequence, $T$, and the probability mass in the first $r$ steps of the lag distribution. They provide a means to quantify the impact of the PSP \sampling primitive and provide stronger convergence guarantees than ASP. Specifically, they do not depend upon the entire lag distribution, which ASP does.

\begin{figure}[h!]
\centering
\includegraphics[scale=0.5]{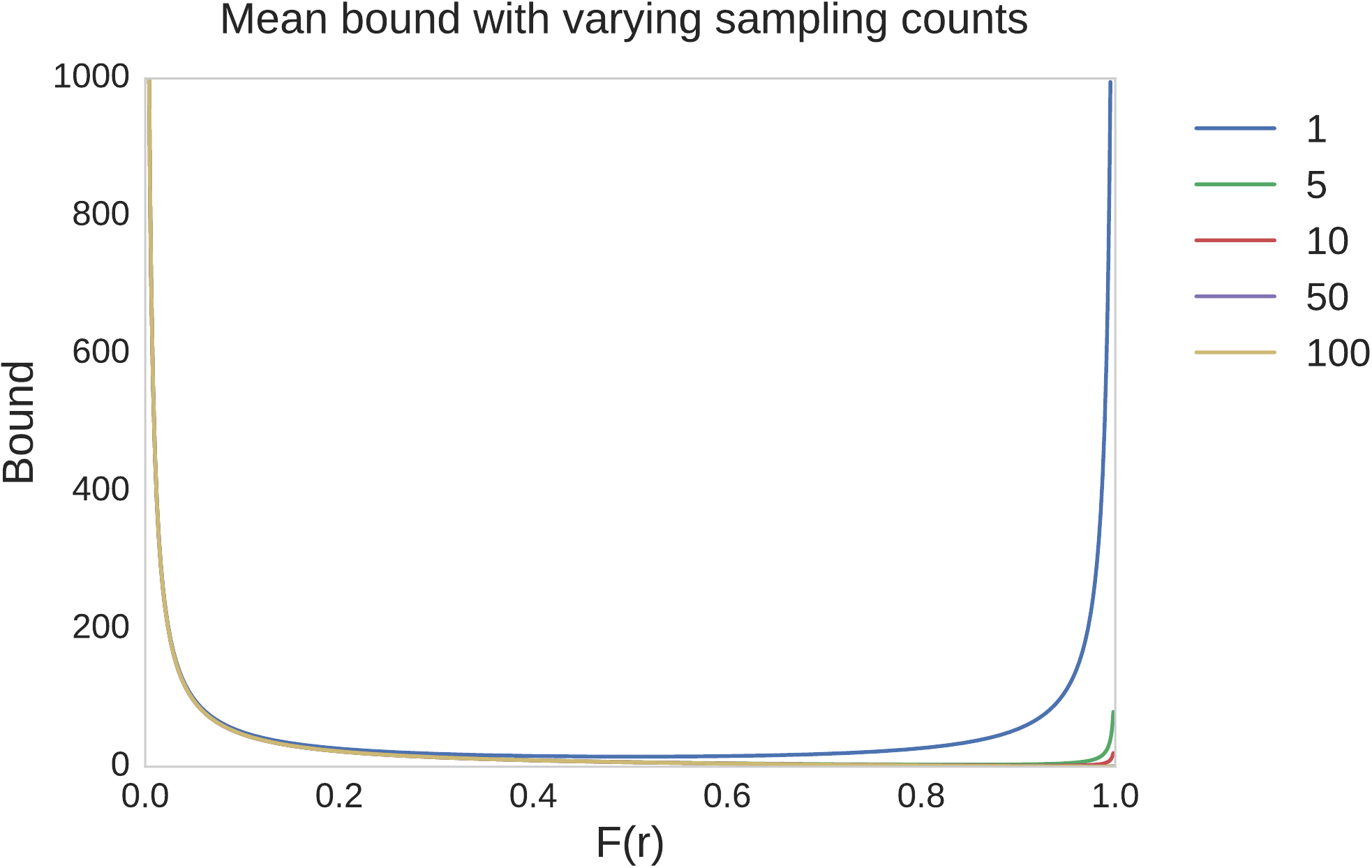}
\caption{Plot showing the bound on the average of the means of the sampling distribution. The sampling count $\beta$ is varied between 1 and 100 and marked with different line colours on the right. The staleness, $r$, is set to 4 with $T$ equal to 10000.}
\label{fig:pspmeanbound}
\end{figure}

\begin{figure}[h!]
\centering
\includegraphics[scale=0.5]{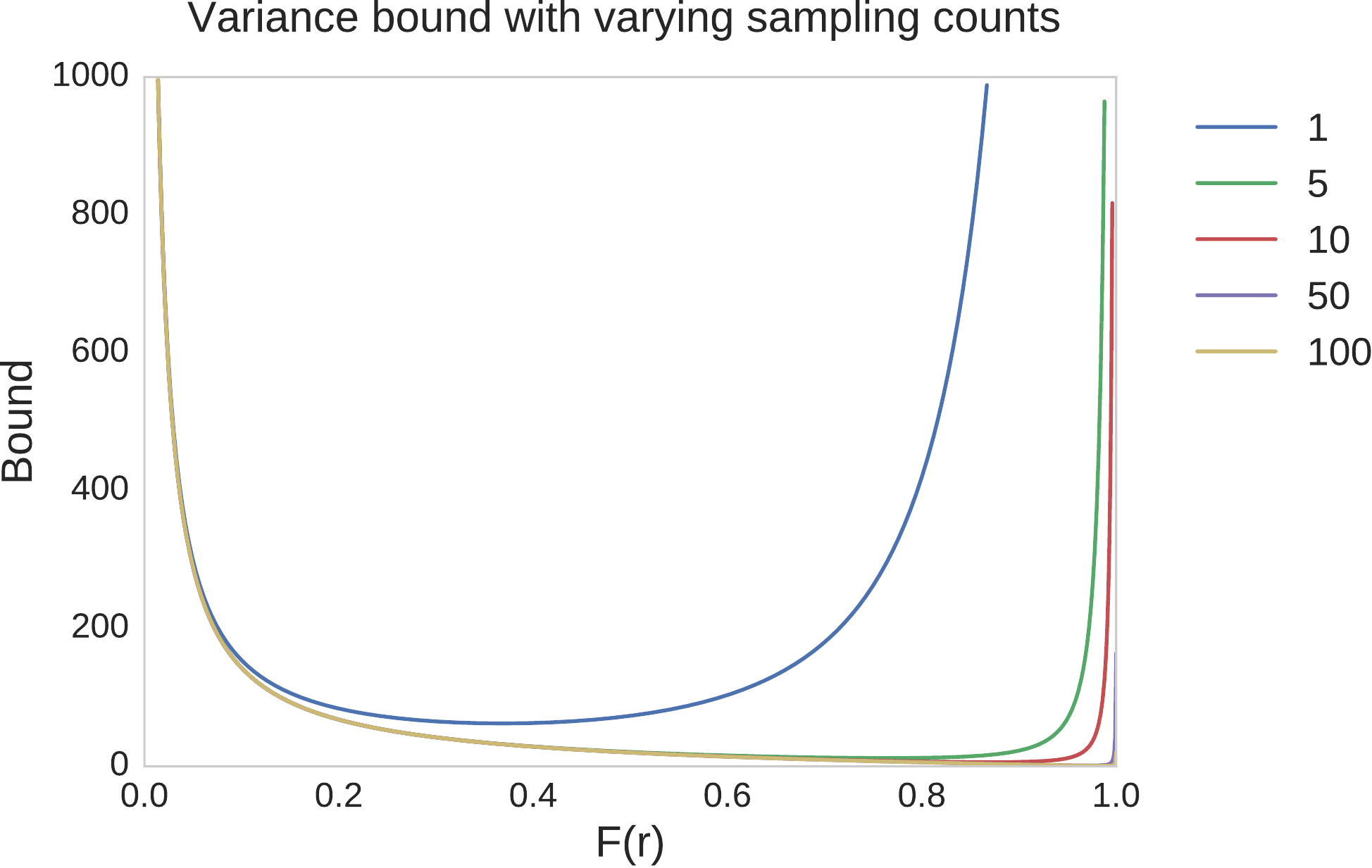}
\caption{Plot showing the bound on the average of the variances of the sampling distribution. The sampling count $\beta$ is varied between 1 and 100 and marked with different line colours on the right. The staleness, $r$, is set to 4 with $T$ equal to 10000.}
\label{fig:pspvariancebound}
\end{figure}

To demonstrate the impact of the \sampling primitive on bounds, Figures \ref{fig:pspmeanbound} and \ref{fig:pspvariancebound} show how increasing the sampling count, $\beta$, (from 1, 5, to 100, marked with different line colours on the right) yields tighter bounds. \textit{Notably, only a small number of nodes need to be sampled to yield bounds close to the optimal. This result has an important implication to justify using sampling primitive in large distributed learning systems due to its effectiveness.}

The discontinuities at $a = 0$ and $a = 1$ reflect edge cases of the behaviour of the barrier method control. Specifically, with $a=0$, no probability mass is in the initial $r$ steps so no progress can be achieved if the system requires $\beta >0$ workers to be within $r$ steps of the fastest worker. If $a = 1$ and $\beta = 0$, then the system is operating in ASP mode so the bounds are expected to be large. However, these are overly generous. Better bounds are $O(T)$ for the mean and $O(T^{2})$ for the variance, which we give in our proof. When $a = 1$ and $\beta \neq 0$, the system should never wait and workers could slip as far as they like as long as they returned to be within $r$ steps before the next sampling point.

\subsection{PSP vs. ASP}

Both PSP and ASP are able to achieve convergence in probability as long as the $\delta$ term in the bounds is relaxed. Provided the algorithms are executed for at least $r+2$ steps ($T > r+1$) and $T > \beta$, then PSP bounds the relaxation of the $\delta$ term independently of the mean of the lag distribution. Furthermore, it only relies on the distribution where the lag is less than $r$. In ASP, the $\delta$ relaxation relies heavily upon the mean of the distribution. Thus, the bound can deteriorate over time if the mean increases and can be far larger than the PSP bound. 

PSP also reduces the impact that the lag distribution can have on the system. For example, PSP can mitigate the impact of long and heavy tailed distributions. Furthermore, PSP is able to provide the same bound even if the lag distribution is changing over time. This is an excellent result for real-world computing environments where fluctuating performance characteristics are typical.

\subsection{PSP vs. SSP}

SSP provides deterministic convergence bounds~\cite{ho2013} due to the guaranteed pre-window updates in the interval $[0,c - s + 1]$. PSP provides no such guarantee. Instead, the SSP update window is replaced by a lag distribution over updates which yields a probabilistic bound on convergence. 

Similar to the comparison to ASP, while SSP can be severely impacted by the stragglers in the system, PSP is much more robust to the heavy-tail distribution regarding working time. More importantly, from system perspective, because PSP does not require global knowledge of the system state as SSP does, the barrier control can be implemented in a fully distributed way. Along with the results presented in Figures \ref{fig:pspmeanbound} and \ref{fig:pspvariancebound}, this indicates that incorporating \sampling primitive can lead to highly scalable data processing and learning system designs.


\section{Conclusion}
\label{sec:conclusion}

In this paper, we proposed a new barrier control technique called Probabilistic Synchronous Parallel. PSP is suitable for data analytic applications deployed in a large and unreliable distributed systems. Comparing to the previous solutions, the proposed one makes a good trade-off between the efficiency and accuracy of iterative learning algorithms. We developed our distributed learning system based on the proposed solution and introduced a new system primitive "\sampling".  We showed that \sampling primitive can be composed with existing barrier controls to derive fully distributed solutions. We evaluated the solution both analytically and experimentally in a realistic setting and our results showed that PSP outperforms existing barrier control solutions and achieve much faster convergence in various settings.


\end{document}